\documentclass[preprint,showpacs,preprintnumbers,amsmath,amssymb,floatfix]{revtex4-1}
\newcommand{\newwidth}{0.675\textwidth}
\newcommand{\newheight}{0.45\textwidth}

\usepackage{graphicx}
\usepackage{dcolumn}
\usepackage{bm,verbatim}

\begin{document}

\title{Transport through an Anderson impurity: Current ringing, non-linear magnetization 
and a direct comparison of continuous-time quantum Monte Carlo and hierarchical quantum master equations} 

\author{R.\ H\"artle$^{1,2}$}
\author{G.\ Cohen$^{2,3}$}
\author{D.\ R.\ Reichman$^{3}$}
\author{A.\ J.\ Millis$^{2}$}
\affiliation{
$^1$ Institut f\"ur theoretische Physik, Georg-August-Universit\"at G\"ottingen, G\"ottingen, D-37077, Germany. \\
$^2$ Department of Physics, Columbia University, New York, NY 10027, USA.  \\
$^3$ Department of Chemistry, Columbia University, New York, NY 10027, USA.  
}

\date{\today}

\begin{abstract}
We give a detailed comparison of the hierarchical quantum master equation
(HQME) method to a continuous-time quantum Monte Carlo (CT-QMC) approach,
assessing the usability of these numerically exact schemes as impurity
solvers in practical nonequilibrium calculations. We review the main
characteristics of the methods and discuss the scaling of the associated
numerical effort. We substantiate our discussion with explicit numerical
results for the nonequilibrium transport properties
of a single-site Anderson impurity. The numerical effort of the HQME
scheme scales linearly with the simulation time but increases (at worst exponentially) 
with decreasing temperature. 
In contrast, CT-QMC is less restricted by temperature at short times, but in general
the cost of going to longer times is also exponential. After establishing
the numerical exactness of the HQME scheme, we use it to elucidate the 
influence of different ways to induce transport through the impurity 
on the initial dynamics, discuss the phenomenon of coherent current oscillations, 
known as current ringing, and explain the non-monotonic temperature dependence of the steady-state
magnetization as a result of competing broadening effects. 
We also elucidate the pronounced non-linear magnetization dynamics, 
which appears on intermediate time scales in the presence of an asymmetric coupling to the electrodes. 
\end{abstract}


\maketitle

\section{Introduction}

Impurity problems are ubiquitous in the theoretical description of
nonequilibrium systems \cite{Andergassen2010,Aoki2013}. They constitute
small entities with a limited number of degrees of freedom that are
coupled to reservoirs with continua of non-interacting degrees of
freedom. One intuitive physical realization 
of such a model is a molecule adsorbed on a surface or contacted by
electrodes \cite{cuevasscheer2010}. A variety of nonequilibrium scenarios may be described in
terms of impurity models, for instance by preparing the impurity in
an excited initial state or by coupling it to different reservoirs in
different thermodynamic states. 
Another intriguing application
occurs in dynamical mean field theory \cite{Georges1996,Freericks2006,Vollhardt2012,Aoki2013},
where lattice problems either in or out of equilibrium are mapped to impurity problems with an environment that 
is determined by a self-consistency criterion. 
This has been important, for instance, in understanding the
metal--insulator transition in materials like transition metal oxides \cite{Georges1996,Nekrasov2005,Vollhardt2012}   
and has become an important paradigm in studying nonequilibrium effects
in extended interacting systems, including thermalization after an
interaction quench \cite{Eckstein2009,Eckstein2013c}, the nonequilibrium
steady state \cite{Pruschke2008,Aron2013} and Bloch oscillations
\cite{Freericks2006,Freericks2008,Eckstein2011} under the influence
of a static electric field. Thus, the
theoretical description of impurity problems is a key element in understanding
a wide range of phenomena, in particular nonequilibrium effects. 

Few exact solutions are available, and a number of methods have been
developed in the past decades to solve nonequilibrium impurity problems.
They can be sorted into two broad categories: approximate 
and numerically 
exact methods. 
Typically, numerically exact methods allow us to simulate some property
of a model in what might be considered a numerical experiment. 
Approximate methods, on the other
hand, may miss important physics or suffer from artifacts due to the
approximations involved. A combination of methods, which operate on different levels 
of approximation 
is often useful and helps to elucidate 
the relevant physical mechanisms 
\cite{Wang2008,Eckel2010,Werner2010b,Schmitt2010,Thoss2011,Hartle2013b,Hartle2014}. 

The nonequilibrium Anderson impurity model has been treated by several
numerically exact methodologies. Some approaches require a discretization
of the electrodes, for example density-matrix renormalization group
\cite{Hassanieh2006,daSilva2008,Kirino2008,HeidrichMeisner2009,Kirino2010,Branschadel2010},
numerical renormalization group \cite{Anders2008,Anders2008b,Schmitt2010}
or multilayer multiconfiguration time-dependent Hartree theory \cite{Thoss2013,Balzer2015}.
These methods are useful at low temperatures and/or voltages. They 
are restricted by revival oscillations and a limited spectral resolution 
of the leads \cite{Zitko2009}. Other approaches
can take advantage of the noninteracting nature of the leads and do
not require discretization. This includes iterative path-integral
schemes \cite{Thorwart2008,Segal2010,Segal2011,Huetzen2012,Weiss2013}, 
which converge only for a limited set of parameters, and stochastic
schemes \cite{Werner2006,Schmidt2008,Werner2009,Schiro2010,Gull2010,Muhlbacher2011,Cohen2013},  
where the growth of the statistical error restricts the
accessible time scales. In the presence of a short memory timescale,
long time scales can be accessed by a combination of reduced dynamics
techniques \cite{Cohen2011} with a short-time numerically exact scheme;
or by the hierarchical master equation method \cite{Zheng2009} where 
the numerical effort scales linearly with the simulation time. The
latter, however, can only be converged if the temperature in the electrodes
is not too low \cite{Hartle2013b,Hartle2014}.

The numerical effort associated with most numerically exact schemes
restricts practical calculations to specific limits \cite{Paaske2005}
or limited ranges of parameters. It is important to delineate the
regimes in parameter space to which each method is applicable, and
in particular to find out where exact results are not available. In
this work, we elucidate the practical limitations of two numerically
exact schemes: the continuous-time quantum Monte Carlo (CT-QMC) method
\cite{Werner2006,Muehlbacher08,Werner2009,Schmidt2009,Schiro2009,Gull2010,Cohen2011,Cohen2013} 
and the hierarchical quantum master equation (HQME) method \cite{Tanimura2006,Welack2006,Jin2008,Hartle2013b,Hartle2014}.
We will discuss the main features of these approaches, characterizing,
in particular, the associated numerical effort. We find that the range
of parameters where the two methods can be applied overlaps, but also
exhibits areas where only one of the methods can be applied. As we
will see, HQME turns out to be the method of choice to study long-term
dynamics if the temperature of the reservoirs is not too low. In contrast,
CT-QMC gives access to the short- and intermediate-time dynamics over 
a wider range of temperatures.

\begin{figure}
\begin{tabular}{p{7cm}lp{4cm}lp{4cm}}
\hspace{1cm}(a) & (b) && (c) & \\
\hspace{0cm}\includegraphics[scale=0.35]{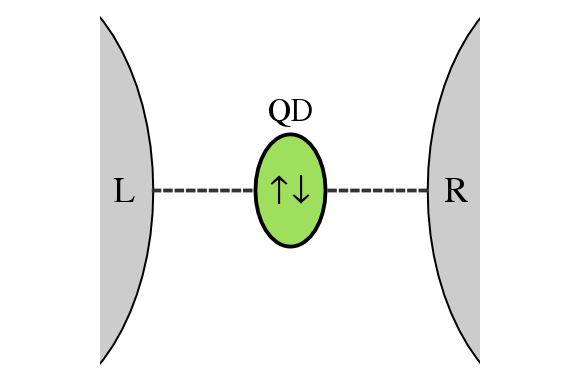} 
&&
&& 
\\[-5.85cm]
&&
\includegraphics[scale=0.4]{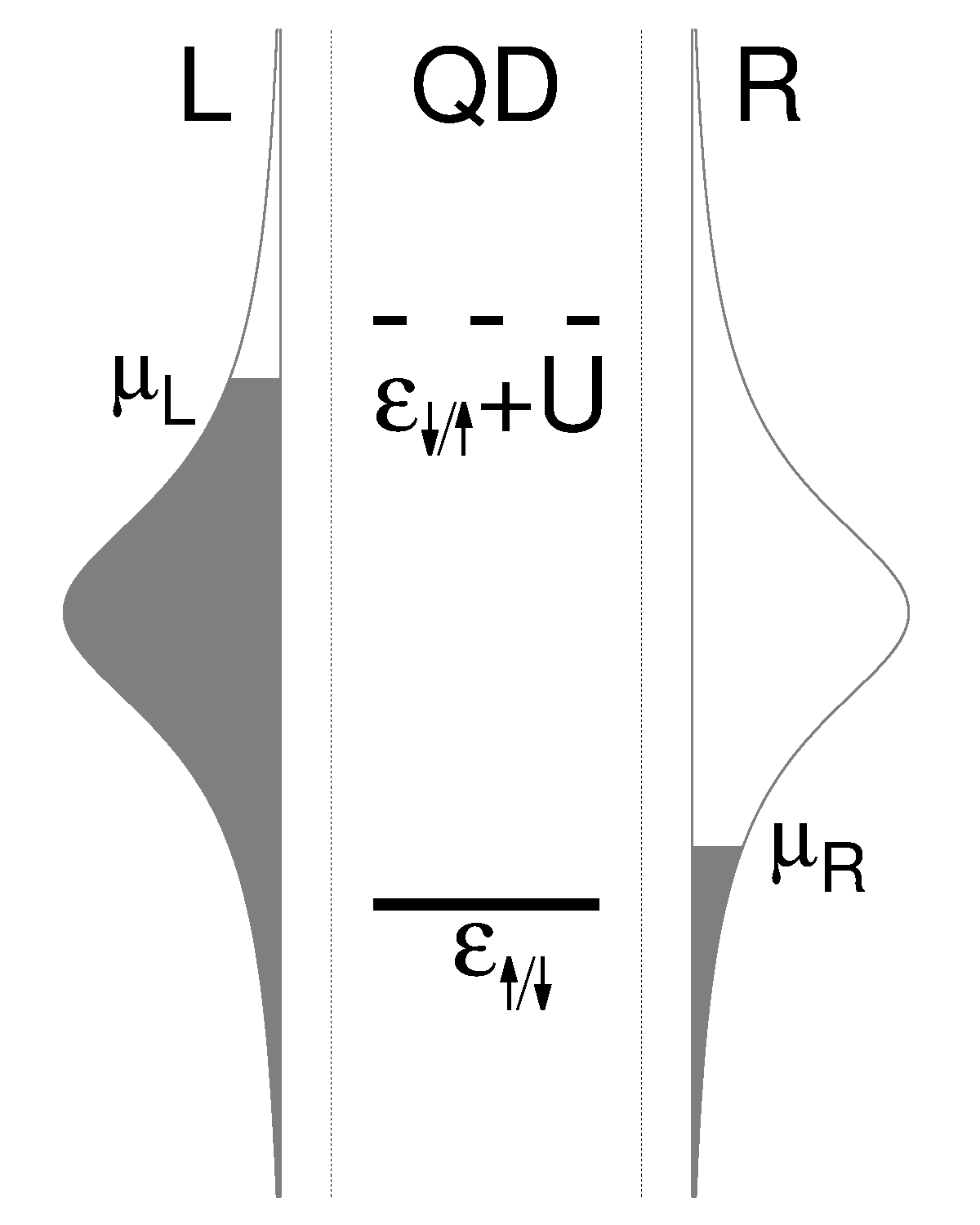}  &&
\\[-5.2cm]
&&&&  
\includegraphics[scale=0.4]{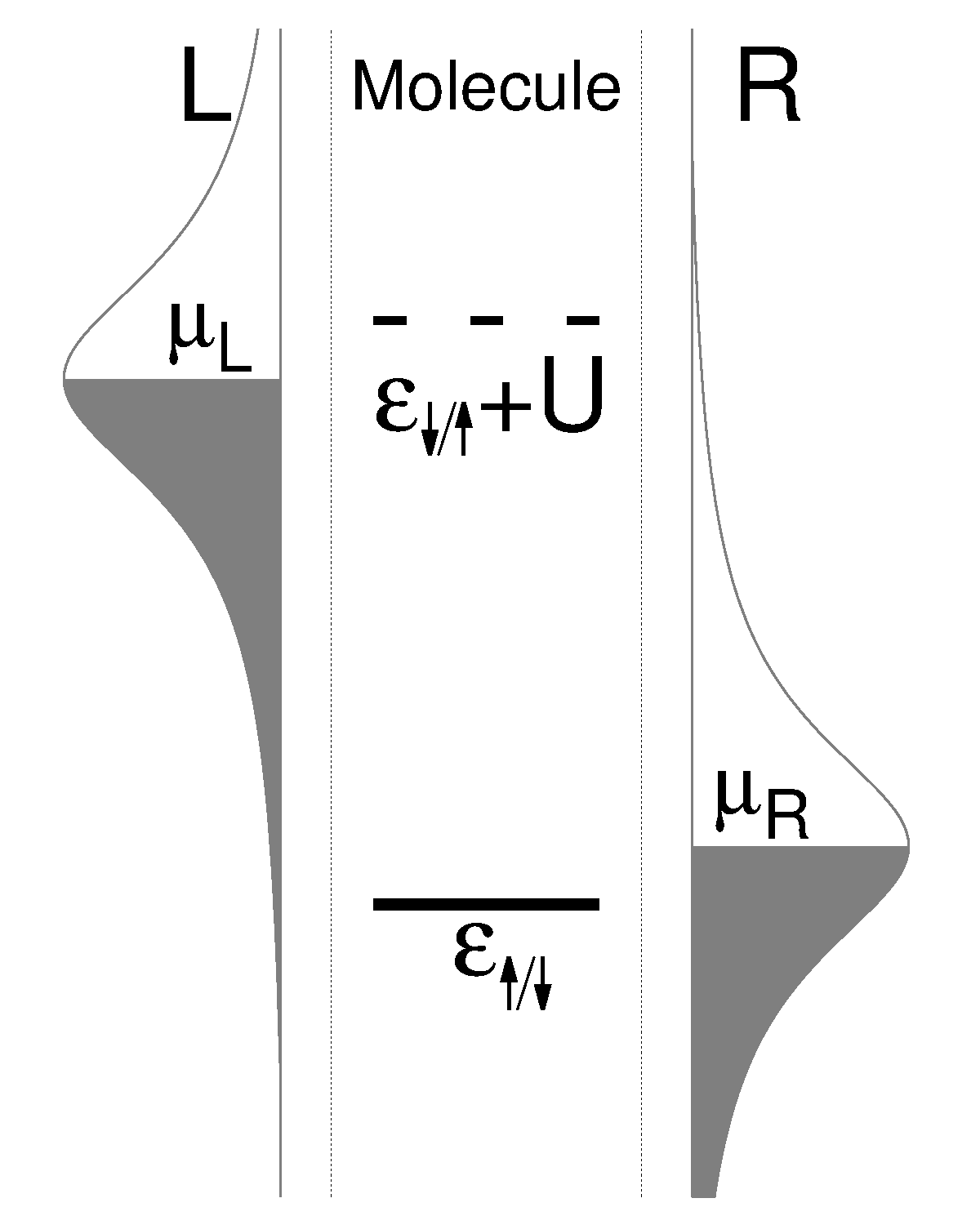}  
\\[0.3cm]
\end{tabular}
\caption{\label{qdfig} Panel (a): Graphical representation of an Anderson impurity, 
which is realized by a quantum dot (QD). The dot is coupled to a left (L) 
and a right electrode (R). Panel (b): Single-particle levels of the quantum dot 
junction depicted in Panel (a). The shaded areas depict the occupied states in the electrodes, 
which, for simplicity, are assumed to provide 
Lorentzian-shaped conduction bands. 
Applying a bias voltage to such a system means that the corresponding chemical 
potentials $\mu_{\text{L/R}}$ are shifted and the electrodes become charged. 
Panel (c): Single-particle levels of a molecular junction. 
In contrast to the quantum dot realization (Panels (a) and (b)), 
the conduction bands are shifted in the same way as the chemical 
potentials. Thus, applying a bias voltage, the electrodes do not become charged. 
}
\end{figure}

We demonstrate our findings using an archetypal nonequilibrium problem:
transport through a single-site Anderson impurity that is coupled
to left and right electrodes (see Fig.\ \ref{qdfig} for a graphical
representation). The most obvious physical realizations of this impurity
problem are quantum dots containing a single spin-degenerate level.
The first such realizations were based on quantum-confinements in patterned 
semiconductor heterostructures \cite{Cronenwett1998,Goldhaber1998}. 
Single-molecule junctions often exhibit similar behavior \cite{Liang02,Pasupathy2004,Osorio2007,Roch2008}, 
but in a setting where experimental techniques give less control over the
parameters of the junction. Additionally, other effects, e.g. due to vibrational degrees of freedom,
are important \cite{Natelson04,Sapmaz05,Pasupathy05,Thijssen06,Parks07,Boehler07,Leon2008,
Repp2009,Ballmann2010,Ballmann2012,Ballmann2013b}. 
In these systems, transport is induced by shifting the electrochemical
potentials in the leads with respect to each other such that electrons
tunnel through the impurity in order to move from the lead with higher
chemical potential to that with lower chemical potential. In semiconductor
heterostructures this shift is achieved by charging or discharging
the leads (\emph{i.e.} by filling or emptying electronic levels; cf.\ Fig.\ \ref{qdfig}b). 
In single-molecule junctions, the leads are less likely to be charged,
and the shift of the electrochemical potentials is accompanied by 
a shift of the respective conduction bands (cf.\ Fig.\ \ref{qdfig}c). 
We will show that these different ways of inducing transport strongly affect the initial dynamics
of the impurity.

We also present exact results for the complex magnetization dynamics of an Anderson impurity 
in various nonequilibrium situations. Typically, this quantity exhibits the slowest 
relaxation behavior \cite{Cohen2013} and, as we will see, exhibits a non-linear 
behavior on all times scales, in particular when an asymmetric coupling 
to the electrodes is considered. 
To date, this dynamics had only been accessible at great 
computational cost using state-to-the-art CT-QMC methods combined
with reduced dynamics \cite{Cohen2013}. The HQME method gives access to exact results of 
this long-lived correlated dynamics and allows us to perfrom a scan over a wide range of 
parameters. It can also be used to derive approximate results and, therefore, to study 
the influence of higher order processes. We are, therefore, able to elucidate the 
origin of the non-monotonic temperature dependence of the magnetization
that was recently reported in Ref.\ \cite{Cohen2013} to be the result of competing broadening effects. 
We also find a pronounced non-linear behavior of the magnetization on intermediate 
(still rather long) time scales (cf.\, \emph{e.g.}, Figs.\ \ref{SymmB} and \ref{ASymmB}). 
In passing we note that the nonequilibrium Anderson impurity
model and its generalizations are of great interest in the field of
strongly correlated materials within the dynamical mean field theory
approximation \cite{Georges1996,Freericks2006,Vollhardt2012,Aoki2013}.

The outline of the article is as follows: In Sec.\ \ref{theorysection}, 
we present the theoretical methodology. This includes a short description
of the single-site Anderson impurity model (Sec.\ \ref{modelhamsec}), 
the HQME method (Sec.\ \ref{hqmesec}) and 
the CT-QMC approach (Sec.\ \ref{ctqmc}).
A discussion of practical aspects of the two methods is given in Sec.\ \ref{hqmevsctqmc}.
Numerical results on the time-dependent transport properties of an
Anderson impurity are presented in Sec.\ \ref{resultssec}, where we first formulate the 
different ways of inducing transport and detail the model parameters (Sec.\  \ref{resultssec}\,A). A direct
comparison of results that are obtained by the HQME method and the CT-QMC approach
is presented in Sec.\ \ref{Comparison}, where we follow 
the time evolution of the electrical current that is flowing through
the impurity, starting from a product initial state where the impurity
is not populated by electrons. These results represent the first explicit
validation that the HQME approach gives numerically exact results. We also explore how the choice
of whether or not to shift the conduction band with the applied bias
voltage affects the results. We then discuss the magnetization dynamics of the
impurity in the presence of an external magnetic field, considering both a symmetric and 
an asymmetric coupling to the electrodes (Sec.\ \ref{magsec}).

\section{Theory}
\label{theorysection}

\subsection{Model Hamiltonian}
\label{modelhamsec}

We study the transport properties of an Anderson impurity that is
coupled to a left (L) and a right (R) electrode or lead. The Hamiltonian
of this well established system,  
\begin{eqnarray}
H & = & H_{\text{imp}}+H_{\text{L}}+H_{\text{R}}+H_{\text{tun}},
\end{eqnarray}
can be decomposed into the impurity Hamiltonian, $H_{\text{imp}}$;
the left and the right lead Hamiltonians, $H_{\text{L}}$ and $H_{\text{R}}$;
and a coupling operator $H_{\text{tun}}$. The impurity Hamiltonian
\begin{eqnarray}
H_{\text{imp}} & = & \sum_{\sigma\in\{\uparrow,\downarrow\}}\epsilon_{\sigma}d_{\sigma}^{\dagger}d_{\sigma}+Ud_{\uparrow}^{\dagger}d_{\uparrow}d_{\downarrow}^{\dagger}d_{\downarrow}.
\end{eqnarray}
represents an electronic level that is addressed by creation and annihilation
operators $d_{\sigma}^{\dagger}$ and $d_{\sigma}$. It can hold a
single spin-up ($\uparrow$) or spin-down ($\downarrow$) electron
at energies $\epsilon_{\uparrow}$ and $\epsilon_{\downarrow}$, respectively,
where $\epsilon_{\uparrow}=\epsilon_{\downarrow}$ without the influence
of an external magnetic field. It can also hold a spin-up and a spin-down
electron simultaneously. Such double occupation costs an additional
charging energy $U>0$, representing repulsive Coulomb interactions.

Each lead is described by a continuum of non-interacting electronic levels 
\begin{eqnarray}
H_{\text{L/R}} & = & \sum_{k\in\text{L/R},\sigma}\epsilon_{\sigma k}c_{\sigma k}^{\dagger}c_{\sigma k}
\end{eqnarray}
with energies $\epsilon_{\sigma k}$. These levels are addressed by
annihilation and creation operators $c_{\sigma k}$ and $c_{\sigma k}^{\dagger}$.
The coupling between the impurity and the electrodes can be characterized
by tunneling matrix elements $V_{\sigma k}$, and the corresponding
coupling operator is written as 
\begin{eqnarray}
H_{\text{tun}} & = & \sum_{k\in\{\text{L+R}\};\sigma}(V_{\sigma k}c_{\sigma k}^{\dagger}d_{\sigma}+\text{h.c.}).
\end{eqnarray}
The resulting tunneling efficiencies, or level-width functions, 
\begin{eqnarray}
\label{tunbare}
\Gamma_{K,\sigma}(\epsilon) & = & 2\pi\sum_{k\in K} \vert V_{\sigma k}\vert^{2} \delta(\epsilon-\epsilon_{\sigma k}),
\end{eqnarray}
depend on the energy of the tunneling electrons ($K\in\text{\ensuremath{\left\{ L,R\right\} }}$).
Throughout this work, we assume that the left and the right electrode
have the same properties, in particular that they have the same temperature $T$. 
The only difference between the electrodes occurs in the presence 
of a bias voltage $\Phi=\mu_{\text{L}}-\mu_{\text{R}}\neq0$. The
position of the chemical potentials of the left and the right lead,
$\mu_{\text{L}}$ and $\mu_{\text{R}}$, are shifted in different
directions. We assume a symmetric voltage drop such that $\mu_{\text{L}}=\Phi/2$
and $\mu_{\text{R}}=-\Phi/2$. The latter is an assumption that, however, is 
not crucial for our discussion.

\subsection{Hierarchical master equation approach} 
\label{hqmesec}

We use two different methods to obtain the transport properties of an Anderson impurity. 
The first of these methods is the HQME approach  
\cite{Tanimura2006,Welack2006,Jin2008,Hartle2013b,Hartle2014}. 
The second one is the CT-QMC method \cite{Werner2006,Werner2009,Schmidt2009,Schiro2009,Cohen2011,Cohen2013}. 
For completeness and to establish the notation, we outline the basics of the HQME method 
in this section. The CT-QMC technique is detailed in the following section, Sec.\ \ref{ctqmc}. 

The central quantity of the HQME technique is the density matrix of
the impurity 
\begin{eqnarray}
 \rho &=& \sum_{l,l'} \rho_{l,l'} \vert \psi_{\text{imp},l} \rangle\langle \psi_{\text{imp},l'} \vert
\end{eqnarray}
where the $\psi_{\text{imp},l}$ represent the corresponding Hilbert space. 
It is obtained by solving the hierarchy of 
equations of motion 
\begin{eqnarray} 
\label{hierarcheom}
\partial_{t}\rho_{j_{1}..j_{\kappa}}^{(\kappa)}(t) & = & -i\left[H_{\text{imp}},\rho_{j_{1}..j_{\kappa}}^{(\kappa)}(t)\right]-\sum_{\lambda\in\{1..\kappa\}}\omega_{K_{\lambda},p_{\lambda}}^{s_{\lambda}}\rho_{j_{1}..j_{\kappa}}^{(\kappa)}(t)\\
 &  & \hspace{-1cm}+\sum_{\lambda\in\{1..\kappa\}}(-1)^{\kappa-\lambda}\eta_{K_{\lambda},\sigma_{\lambda},p_{\lambda}}^{s_{\lambda}}d_{m_{\lambda}}^{s_{\lambda}}\rho_{j_{1}..j_{\kappa}/j_{\lambda}}^{(\kappa-1)}(t)+\sum_{\lambda\in\{1..\kappa\}}(-1)^{\lambda}\eta_{K_{\lambda},\sigma_{\lambda},p_{\lambda}}^{\overline{s}_{\lambda},*}\rho_{j_{1}..j_{\kappa}/j_{\lambda}}^{(\kappa-1)}(t)d_{m_{\lambda}}^{s_{\lambda}}\nonumber \\
 &  & \hspace{-1cm}-\sum_{j_{\kappa+1},\sigma_{\kappa+1}}\left(d_{\sigma_{\kappa+1}}^{\overline{s}_{\kappa+1}}\rho_{j_{1}..j_{\kappa}j_{\kappa+1}}^{(\kappa+1)}(t)-(-1)^{\kappa}\rho_{j_{1}..j_{\kappa}j_{\kappa+1}}^{(\kappa+1)}(t)d_{\sigma_{\kappa+1}}^{\overline{s}_{\kappa+1}}\right).\nonumber 
\end{eqnarray}
A detailed derivation of these equations can be found in Refs.\ \cite{Jin2008,Hartle2013b}.
The density matrix of the impurity enters at the $0^{\mathrm{th}}$
tier as $\rho^{(0)}(t)=\rho(t)$. The auxiliary operators $\rho_{j_{1}..j_{\kappa}}^{(\kappa)}(t)$
encode the dynamics of the impurity that originates from the coupling
to the electrodes, starting from a product initial state or, equivalently,
$\rho_{j_{1}..j_{\kappa}}^{(\kappa)}(0)=0$ ($\kappa>1$). They are
distinguished by superindices $j_{\lambda}=(K,\sigma,s,p)$, which
involve a lead index $K$, a spin index $\sigma$ and an index $s\in\{+,-\}$ that
corresponds to the creation and annihilation of electrons.
The index $p$ is related to a decomposition of the lead correlation
functions 
\begin{eqnarray}
C_{K,\sigma}^{s}(t) & = & \int_{-\infty}^{\infty}\frac{\text{d}\omega}{2\pi}\,\text{e}^{si\omega t}\Gamma_{K,\sigma}(\omega)f_{K}^{s}(\omega)\label{corrfunctions}\\
 & = & \sum_{p}\eta_{K,\sigma,p}^{s}\text{e}^{-\omega_{K,p}^{s}t}
\end{eqnarray}
by a set of exponential functions, $\text{e}^{-\omega_{K,p}^{s}t}$,
where we use the short-hand notations $f_{K}^{+}(\omega)=f_{K}(\omega)$
and $f_{K}^{-}(\omega)=1-f_{K}(\omega)$ with $f_{K}(\omega)$ representing
the Fermi distribution function of lead $K$. The use of exponential functions facilitates 
a systematic closure of the hierarchy (\ref{hierarcheom}) \cite{Jin2008,Hartle2013b}. 
We obtain the frequencies
$\omega_{K,p}^{s}$ and the amplitudes $\eta_{K,\sigma,p}^{s}$, using
a Pade decomposition scheme \cite{Hu2010,Hu2011}. 
Explicit expressions can be found in Refs.\ \onlinecite{Hartle2013b,Hartle2014}.

In principle, the solution of the full hierarchy (\ref{hierarcheom})
is exact. In practical calculations, however, the number of Pade poles 
that can be included is limited. For the present studies, we obtained converged results using 
$100$ Pade poles. In addition, the hierarchy of equations
of motion (\ref{hierarcheom}) needs to be truncated. To this end,
we estimate the importance of the operators $\rho_{j_{1}..j_{\kappa}}^{(\kappa)}(t)$
by assigning them the following importance value: \cite{Hartle2013b} 
\begin{eqnarray}
\left\vert \left(\prod_{\lambda=1..\kappa}\frac{1}{\sum_{\lambda'=1..\lambda}\text{Re}[\omega_{K_{\lambda'},p_{\lambda'}}^{s_{\lambda'}}]}\right)\cdot\left(\prod_{\lambda=1..\kappa}\frac{\eta_{K_{\lambda},\sigma_{\lambda},p_{\lambda}}^{s_{\lambda}}}{\text{Re}[\omega_{K_{\lambda},p_{\lambda}}^{s_{\lambda}}]}\right)\right\vert .\label{ampli}
\end{eqnarray}
We include only those operators in our calculations which have a value
larger than a certain threshold value $A_{\text{th}}$. In addition,
all operators of the zeroth and first tier are taken into account
regardless of their assigned importance value. The truncation allows us to reduce
the numerical effort to a practical level. While this procedure represents
an approximation, exact results can be obtained by systematically
reducing the threshold value $A_{\text{th}}$ until the results converge
to within the desired precision. As the importance criterion (\ref{ampli}) 
involves the ratio between the amplitudes $\eta_{K,\sigma,p}^{s}$ 
and the frequencies $\omega_{K,p}^{s}$, that is, effectively the ratio $\Gamma_{K,\sigma}/T$, 
convergence can be achieved more easily at higher temperatures, and 
the numerical effort increases substantially at lower temperatures. 
We will elaborate on this statement in Sec.\ \ref{Comparison}, where
we show that ``large enough'' means in the present context that
the temperature should be above the Kondo temperature. An example
of a convergence analysis is given in the appendix.

A central characteristic of the technique is that the equations of
motion (\ref{hierarcheom}) are local in time. Thus, the numerical
effort of computing the time-dependent operators $\rho_{j_{1}..j_{\kappa}}^{(\kappa)}(t)$
scales linearly with the simulation time $t$. All of our numerical evidence 
(cf., \emph{e.g.}, the appendix) shows that the quality of the associated 
effective expansion of the time evolution operator is independent of the simulation time $t$. 
We can therefore conclude that the numerical effort of the HQME scheme scales 
linearly with the simulation time. This allows us to access 
the nonequilibrium dynamics of an interacting impurity system on extremely
long time scales (see, \emph{e.g.}, Ref.\ \cite{Hartle2014}, where
we simulated the dynamics of an interacting quantum dot system up
to $t\sim10^{4}/\Gamma$). The price of this is a large number of
unknown auxiliary operators $\rho_{j_{1}..j_{\kappa}}^{(\kappa)}(t)$
that need to be determined. At a given time $t$, they encode the
history of the interplay between the impurity and the electrode at
earlier times, and contain all the information necessary to continue
propagating the density matrix to the next time step. 


In addition to the importance criterion (\ref{ampli}), the hierarchy
can be truncated at a specific tier $\tilde{\kappa}$. This corresponds
effectively to a hybridization expansion of the time evolution operator
of the density matrix $\rho(t)$ that is valid up to 
$\mathcal{O}(\Gamma^{\tilde{\kappa}_{K,\sigma}}/\text{min}(D,k_{\text{B}}T)^{\tilde{\kappa}})$
\cite{Hartle2013b}. Such a truncation is not exact but facilitates
a perturbative analysis, which, in principle, can be driven to arbitrary
order. We can therefore assess the importance of each tier/order by
comparison to the exact converged results (see Sec.\ \ref{magsec}).
In this context, it should be noted that the hierarchy (\ref{hierarcheom})
terminates automatically at the second tier for $U\rightarrow0$ \cite{Jin2008,Jin2010}.
In the appendix, we demonstrate the convergence of our approach to
the exact $U=0$ result.

\subsection{Continuous-time quantum Monte Carlo approach}
\label{ctqmc}

In order to establish the numerical exactness of the HQME approach,
it is useful to compare it to another numerically exact approach based 
on entirely different principles. 
As noted in the introduction, a wide variety of numerically exact
methods with various advantages and limitations has been applied to
nonequilibrium impurity models 
\cite{Anders2005,Anders2008,Thorwart2008,HeidrichMeisner2009,Wang09,Eckel2010,Thoss2011,Segal2010,Gull2010,Gull2011,Simine13,Wilner2013,Wolf2014,WangCohen2014}. 
We have chosen to compare our results with those of a continuous-time
quantum Monte Carlo method \cite{Gull2011,Cohen2014}. CT-QMC
algorithms are capable of solving a variety of impurity models by
stochastically summing all terms in an exact diagrammatic expansion
around some analytically solvable limit. 

Dynamics and nonequilibrium require a real time (rather than
an equilibrium imaginary time) formulation of the method to conserve
numerical exactness. The first real-time implementations addressed vibrations
in junctions using hybridization expansions \cite{Muehlbacher08,Schiro2009}, 
with subsequent work treating the Anderson impurity model \cite{Werner2009}
and developing expansion in the interaction \cite{Werner2010b}.
This first generation of methods was mostly suitable for accessing
very short times or weakly interacting systems. A much wider range
of parameters and timescales is accessible to bold-line algorithms \cite{Gull2010b,Gull2011}, 
which begin from a diagrammatic approximation containing an infinite
subset of diagrams corresponding to a low-order self energy, and sum
all corrections to it in terms of renormalized skeleton diagrams.
These methods can be further augmented by reduced dynamics techniques,
which give access to essentially any timescale in cases where the
system exhibits a short memory timescale \cite{Cohen2011,Cohen2013,Cohen2013b}. 
More recently, these techniques were extended from single-time properties
to correlation functions in equilibrium \cite{Cohen2014}
and nonequilibrium \cite{Cohen2014b}. 

In this work, we compare our HQME results to bold-line CT-QMC formulated
around the one-crossing approximation (OCA) \cite{Eckstein2010,Ruegg2012}. OCA is 
a strong-coupling approximation. It represents an extension of the non-crossing
approximation and generally performs well near half-filling and outside
the Kondo regime. Convergence becomes easier when the OCA is more
accurate, but all the CT-QMC data presented here has been converged
up to times $t=2/\Gamma$, and can therefore 
be assumed to be numerically exact, independently of the OCA. A detailed
technical discussion of the method can be found in Ref.~\onlinecite{Cohen2014}.

While no significant problems occur up to $t=2/\Gamma$, we note that
in general it can be difficult to obtain converged CT-QMC data at
long times (and in the absence of a short memory), since the sign-problem 
results in an exponential growth of the statistical error with time. Bold-line algorithms significantly
improve the performance of these algorithms, but do not eliminate
this problem. ``Boldification'' additionally depends on one's ability
to solve the underlying self-consistent diagrammatic approximation
(OCA in this case). This generally implies an initial step with its
own computational and memory demands, both of which increase polynomially
with the simulated time. Higher-order self-energies reduce the sign-problem
in the CT-QMC step, but the cost of the diagrammatic approximation
often becomes prohibitive \cite{Eckstein2010}. 

A particularly simple example, which illustrates how this polynomial
scaling can become a bottleneck, is when several energy scales which
are orders of magnitude apart are present in the problem. Unless an
efficient multi-scale representation of the data is possible, the
diagrammatic procedure -- which is implemented on a discrete lattice -- suffers
from the need to use very small time steps in the discretization.
The effort involved in solving the self-consistent equations is then
polynomial in the number of time steps, and therefore grows very rapidly
with simulation time. Importantly, this never occurs with the time-local 
HQME, where the computational effort is inherently \emph{linear} in
time.

\subsection{Advantages and Drawbacks: When to use which method}
\label{hqmevsctqmc}

The HQME and CT-QMC schemes are similar in the sense that they are
both based on a hybridization expansion. The methods differ in the
way the expansion is carried out. For the HQME approach, we expand
the time evolution operator of the reduced density matrix. If the
expansion converges, one obtains exact results. If not, one obtains
only approximate results, even at short simulation times. In contrast,
the CT-QMC approach represents a stochastic sum over all possible
trajectories that the system may follow during the simulation time. 
The statistical error or, equivalently, the numerical effort
increases in the same way as the number of relevant trajectories increases
with the simulation time \cite{Gull2010}. 
The number of relevant trajectories grows exponentially with 
the simulation time, so QMC methods work well for short times, 
but long-lived correlated dynamics
is often out of the method's reach \cite{Cohen2013}. HQME, in contrast,
gives access to long-lived dynamics, because the associated numerical
effort scales linearly with the simulation time and because all of our numerical evidence 
points out that the quality of the associated expansion is independent of the simulation time $t$. 
This is, on one hand, evident from Eqs.\ (\ref{hierarcheom}) and, on the other hand,  
explicitly demonstrated in Sec.\ \ref{Comparison}, the appendix or, for example, 
in Ref.\ \onlinecite{Hartle2014}. 
We also note in passing that because 
both HQME and CT-QMC are formulated in continous time, 
they are free from discretization (non-zero time step) errors, 
so that very short times may easily be studied.

We further discuss the numerical effort of the HQME method. Apart
from the linear scaling with the simulation time, it depends on the
specific problem, in particular the number $N$ of distinct superindices
$j_{\lambda}$. The latter is given by the complexity of the correlation
functions (\ref{corrfunctions}). The number of auxiliary operators
scales as $N^{\tilde{\kappa}}/\tilde{\kappa}!$, assuming that the
hierarchy (\ref{hierarcheom}) is truncated at the $\tilde{\kappa}$th
tier. The importance criterion (\ref{ampli}) further reduces the
numerical effort to $N^{\tilde{\kappa}-1}/\tilde{\kappa}!$, cutting
out a hypersurface of the total index space \cite{Hartle2014}. The
criterion also demonstrates that fewer terms are needed at higher
temperatures (cf.\ the discussion following Eq.\ \ref{ampli} in Sec.\ \ref{hqmesec} or Ref.\ \onlinecite{Hartle2013b}). 
In addition, each auxiliary operator involves a number
of coefficients that is given by the dimension of the Hilbert space of the impurity.
In general, the size of this space results in an exponential scaling
of the numerical effort with the spin or orbital degrees of freedom
of the impurity. 
In many cases, however, one is interested in 
single-particle quantities or one can restrict the attention to an active space of 
considerably reduced dimension, possibly enabling a power-law scaling. 
An explicit demonstration of this conjecture will be subject of future research. 

\begin{figure}
\resizebox{\newwidth}{\newheight}{
\includegraphics{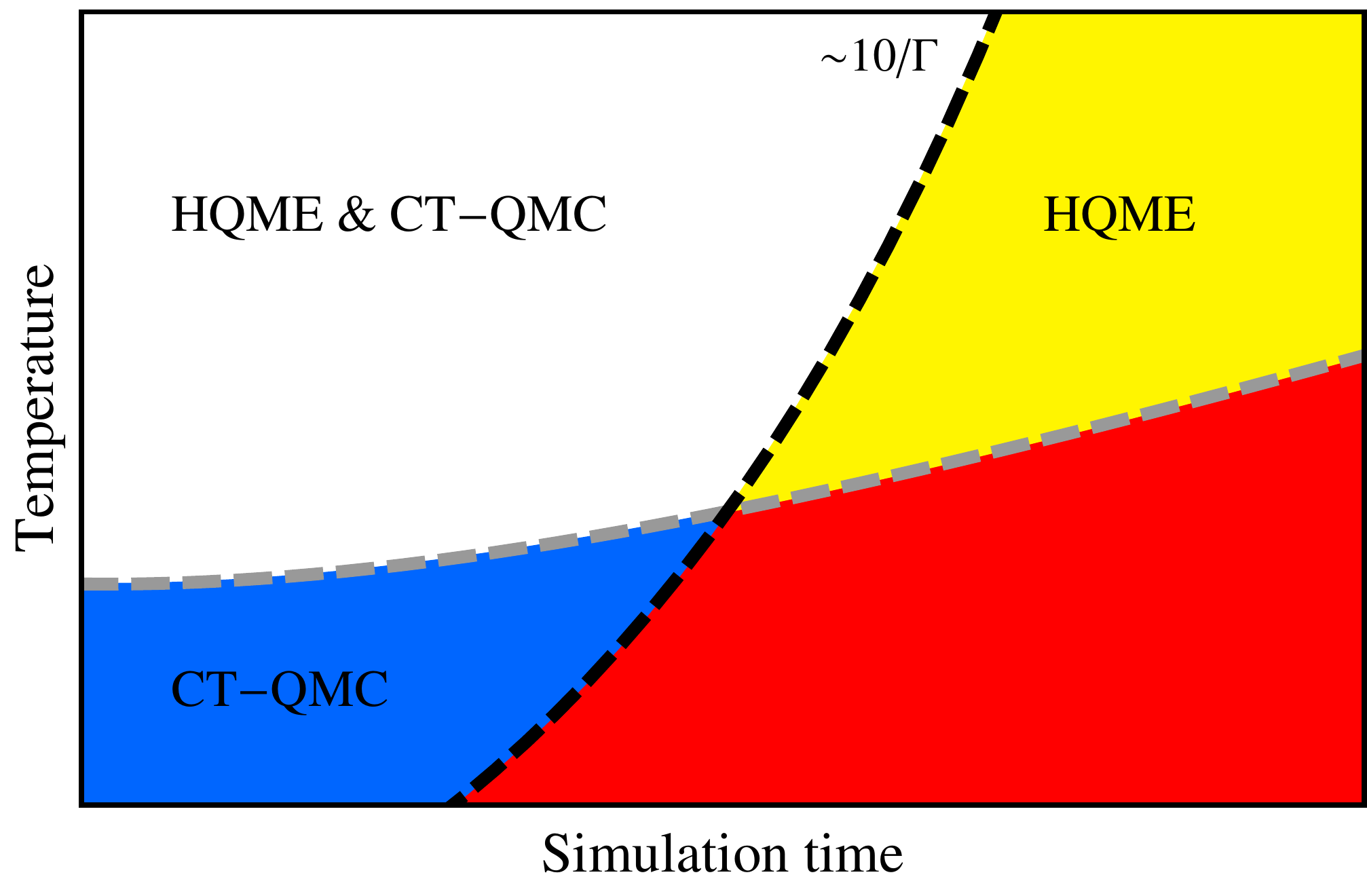}
} 
\caption{(Color online)\label{usefulness} 
Sketch of the areas in simulation
time and temperature where HQME and CT-QMC are useful. The dashed
lines represent the exponential growth of the numerical effort with
the simulation time in CT-QMC (black line) and with the number of coefficients
in HQME (gray line) that increases (at worst exponentially) with
the inverse temperature. High temperatures and short simulation times
are accessible by both methods (white area). Very long simulation
times are accessible only by HQME (yellow area). The low temperature
regime is reserved for CT-QMC (blue area). At low temperatures and
if long simulation times are required, both CT-QMC and HQME cannot
be used. For the given problem, the exponential walls are located 
around the Kondo temperature for HQME and time scales $\sim10/\Gamma$ 
for CT-QMC (cf.\ Sec.\ \ref{Comparison} and Refs.\ \onlinecite{Cohen2011,Cohen2013,Cohen2014}).}
\end{figure}

We summarize our discussion regarding the validity and usefulness
of the methods in Fig.\ \ref{usefulness}. HQME and CT-QMC have common
regimes where they give the same result (high temperature, short times).
This will be demonstrated in Sec.\ \ref{Comparison} explicitly.
There are also regimes where only one of the methods can
be used in practice. Low temperature systems requiring long simulation
times cannot be probed by either of the methods, unless
they also exhibit a short memory timescale, in which case reduced
dynamics techniques may be applicable. This means that slow dynamics
deep in the Kondo regime remain largely inaccessible for both methods. 
The dashed lines in Fig.\ \ref{usefulness} represent the exponential wall 
that is hit in CT-QMC with an increasing simulation time and in HQME with the number 
of operators that needs to be taken into account at decreasing temperatures. 
These walls are ``soft'' in the sense that these boundaries can be pushed by 
more efficient codes and procedures, more powerful computer architectures 
and merely a larger investment of CPU time. They also depend to a large extent on the specific problem. 
Therefore, we refrain from putting specific numbers at this point, but will elaborate 
on the boundaries specific for the Anderson impurity model in the strong coupling regime ($U/(\pi\Gamma)>1$) below.

\subsection{Observables of interest}

We characterize the nonequilibrium transport properties of an Anderson
impurity by its magnetization $m$ and the electrical current $I$
that is flowing through the impurity in the presence of a bias voltage.
The magnetization is given by the diagonal elements of the impurity 
density matrix 
\begin{eqnarray}
m(t) & = & \rho_{\downarrow,\downarrow}(t)-\rho_{\uparrow,\uparrow}(t),
\end{eqnarray}
where we use the basis $\left\{ \vert00\rangle,\vert\uparrow\rangle,\vert\downarrow\rangle,\vert\uparrow\downarrow\rangle\right\} $.
This basis includes the states of the impurity with no electron, $\vert00\rangle$;
a single spin-up electron, $\vert\uparrow\rangle$; a single spin-down
electron, $\vert\downarrow\rangle$; and two electrons, $\vert\uparrow\downarrow\rangle$.

The electrical current flowing through the impurity is 
related to the charge flow in and out of each lead K: 
\begin{eqnarray}
I_{K} & = & -e\frac{\text{d}}{\text{d}t}\sum_{k\in K}\langle c_{k}^{\dagger}c_{k}\rangle,\label{firstcurrent}
\end{eqnarray}
where $-e$ denotes the charge of an electron. Using the auxiliary
operators $\rho_{j}^{(1)}(t)$, it can be written as \cite{Jin2008,Hartle2013b}
\begin{eqnarray}
I_{K}(t) & = & e\sum_{K,\sigma,p}\left(\text{Tr}_{\text{imp}}\left[\rho_{K,\sigma,+,p}^{(1)}(t)d_{\sigma}\right] - 
\text{Tr}_{\text{imp}}\left[d_{\sigma}^{\dagger}\rho_{K,\sigma,-,p}^{(1)}(t)\right]\right).
\end{eqnarray}

\section{Results}
\label{resultssec}

\subsection{Formulation of the transport problem}

In the following, we investigate transport and relaxation phenomena
of a charge-symmetric Anderson impurity where $\epsilon_{\uparrow}+\epsilon_{\downarrow}=-U$
(see Figs.\ \ref{qdfig}b and \ref{qdfig}c). We follow the time evolution
from a product initial state where the impurity is not correlated
with the electrodes and carries no electron (\emph{i.e.} $\rho_{00,00}(t=0)=1$
while all other elements of the reduced density matrix are zero).
We focus on the intermediate to strong coupling regime, choosing $U=8\Gamma$
(or, equivalently, $\frac{U}{\pi\Gamma}\approx2.5$), where $\Gamma$
denotes the hybridization strength between the impurity and the electrodes
at the Fermi level. We have chosen this regime because it 
represents the most challenging regime for the HQME framework and indeed for most theoretical
treatments (since simple approximations generally work for either
$\frac{U}{\Gamma}\rightarrow0$ or $\frac{U}{\Gamma}\rightarrow\infty$),
while also exhibiting a rich and interesting variety of nonequilibrium
phenomena. 

We take the tunneling efficiencies (Eq.\ (\ref{tunbare})) to be 
\begin{eqnarray}
\Gamma_{\text{L/R},\sigma}(\epsilon)=\Gamma\alpha_{\text{L/R}}\frac{D^{2}}{(\epsilon-S\mu_{\text{L/R}})^{2}+D^{2}},
\end{eqnarray}
where we assume Lorentzian-shaped conduction bands in the electrodes
with $D=10\Gamma$. Note that the shape of the conduction bands is
not crucial for our discussion but beneficial for the numerical evaluation
of the HQME. The parameters $\alpha_{\text{L/R}}$ are either $\alpha_{\text{L}}=1=\alpha_{\text{R}}$
corresponding to a symmetric coupling of the impurity to the electrodes
or $\alpha_{\text{L}}=1=4\alpha_{\text{R}}$ to simulate an asymmetric
impurity-electrode coupling. The parameter $S$ is used to control
whether the conduction bands are shifted with the applied bias voltage
($S=1$; cf.\ Fig.\ \ref{qdfig}c) or not ($S=0$; cf.\ Fig.\ \ref{qdfig}b). 
While the former situation corresponds to
a scenario that is typically found, for example, in transport through
a single-molecule junction, the latter is often used to describe transport
through quantum dot structures that are based on semiconductor heterostructures.

Our analysis includes two parts. In the first part, Sec.\ \ref{Comparison},
we present results for the electrical current that is flowing through
the impurity in the presence of a bias voltage. Thereby, we give a
detailed comparison of HQME and CT-QMC results, which, on one hand,
validates the HQME framework that we introduced in Ref.\ \cite{Hartle2013b}
(and outlined in Sec.\ \ref{hqmesec}) and, on the other hand, allows
us to explore the different initial dynamics of the electrical current
with respect to the two values of the shift parameter $S$. Second, in Sec.\ \ref{magsec},
we focus on the dynamics of the dot observable that takes longest
to approach its steady state value: the magnetization $m$. We
simulate the effect of a magnetic field by shifting the spin-up and
the spin-down level of the impurity with the field strength $h$, 
\begin{eqnarray}
\label{effectofmagneticfield}
\epsilon_{\uparrow} & = & -\frac{U}{2}+h,\\
\epsilon_{\downarrow} & = & -\frac{U}{2}-h,
\end{eqnarray}
and study the evolution of $m$ to its field-dependent steady state values. 
This allows us to demonstrate that HQME gives access to long-lived
correlated dynamics that, to date, had only been accessible at great
computational cost using state-to-the-art CT-QMC methods combined
with reduced dynamics \cite{Cohen2013}. Moreover, we elucidate the
origin of the non-monotonic temperature dependence of the magnetization
that was recently reported in Ref.\ \cite{Cohen2013}. All model
parameters are listed in Tab.\ \ref{parameters}, where the different
parameter sets are labeled by the figure depicting the corresponding
results.

\begin{table}
\begin{center}
\begin{tabular}{|*{11}{ccc|}}
\hline \hline 
&Fig.\ &&& $\epsilon_{\uparrow}$&&&$\epsilon_{\downarrow}$&&&$U$&&& 
$\alpha_{\text{L}}$&&& $\alpha_{\text{R}}$&&& $S$&&&$h$&&&
$k_{\text{B}}T$&&&$D$&\\ \hline 
& \ref{CompHigh}  &&& -4 &&& -4 &&& 8 &&& 1 &&& 1             &&& 0 &&& 0 &&& 5 &&& 10 &\\
& \ref{CompInter} &&& -4 &&& -4 &&& 8 &&& 1 &&& 1             &&& 0 &&& 0 &&& 1 &&& 10 &\\
& \ref{CompLow}   &&& -4 &&& -4 &&& 8 &&& 1 &&& 1             &&& 0 &&& 0 &&& $\frac{1}{5}$ &&& 10 &\\
& \ref{CompShift} &&& -4 &&& -4 &&& 8 &&& 1 &&& 1             &&& 1 &&& 0 &&& 1 &&& 10 &\\
& \ref{SymmB}     &&& -4 &&& -4 &&& 8 &&& 1 &&& 1             &&& 0 &&& 2 &&& $\frac{1}{2}$ -- 10 &&& 10 &\\
& \ref{mtfig}     &&& -4 &&& -4 &&& 8 &&& 1 &&& 1             &&& 0 &&& 2 &&& $\frac{1}{2}$ -- 10 &&& 10 &\\
& \ref{ASymmB}    &&& -4 &&& -4 &&& 8 &&& 1 &&& $\frac{1}{4}$ &&& 0 &&& 2 &&& $\frac{1}{2}$ -- 10 &&& 10 &\\
& \ref{convfig}a  &&& -4 &&& -4 &&& 8 &&& 1 &&& 1             &&& 0 &&& 0 &&& 1 &&& 10 &\\
& \ref{convfig}b  &&& -4 &&& -4 &&& 0 &&& 1 &&& 1             &&& 0 &&& 0 &&& 1 &&& 10 &\\
\hline \hline
\end{tabular}
\end{center}
\caption{\label{parameters}
Parameters for the quantum dot devices that are investigated in this article. 
All energy values are given in units of the hybridization strength $\Gamma$.  
}
\end{table}

\subsection{Time-dependent electrical current: Comparison of HQME and CT-QMC}
\label{Comparison}

The primary goal of this section is to provide the first direct comparison
of results that have been obtained by the HQME and CT-QMC approach.
We thus validate the HQME scheme with respect to an established method
and, at the same time, demonstrate explicitly that our truncation
scheme is consistent and gives numerically exact results once
convergence is achieved. As the importance criterion (\ref{ampli})
suggests, the HQME expansion works best for high temperatures. Accordingly,
we start our comparison at a relatively high temperature in the electrodes,
$\beta=(k_{\text{B}}T)^{-1}=0.2/\Gamma$, and continue with an intermediate,
$\beta=1/\Gamma$, and a low temperature, $\beta=5/\Gamma$. The latter is close to 
the Kondo temperature of our setup, $\beta_{\text{Kondo}}\approx15/\Gamma$
\cite{Wiegmann1982}.

\begin{figure}
\resizebox{\newwidth}{\newheight}{
\includegraphics{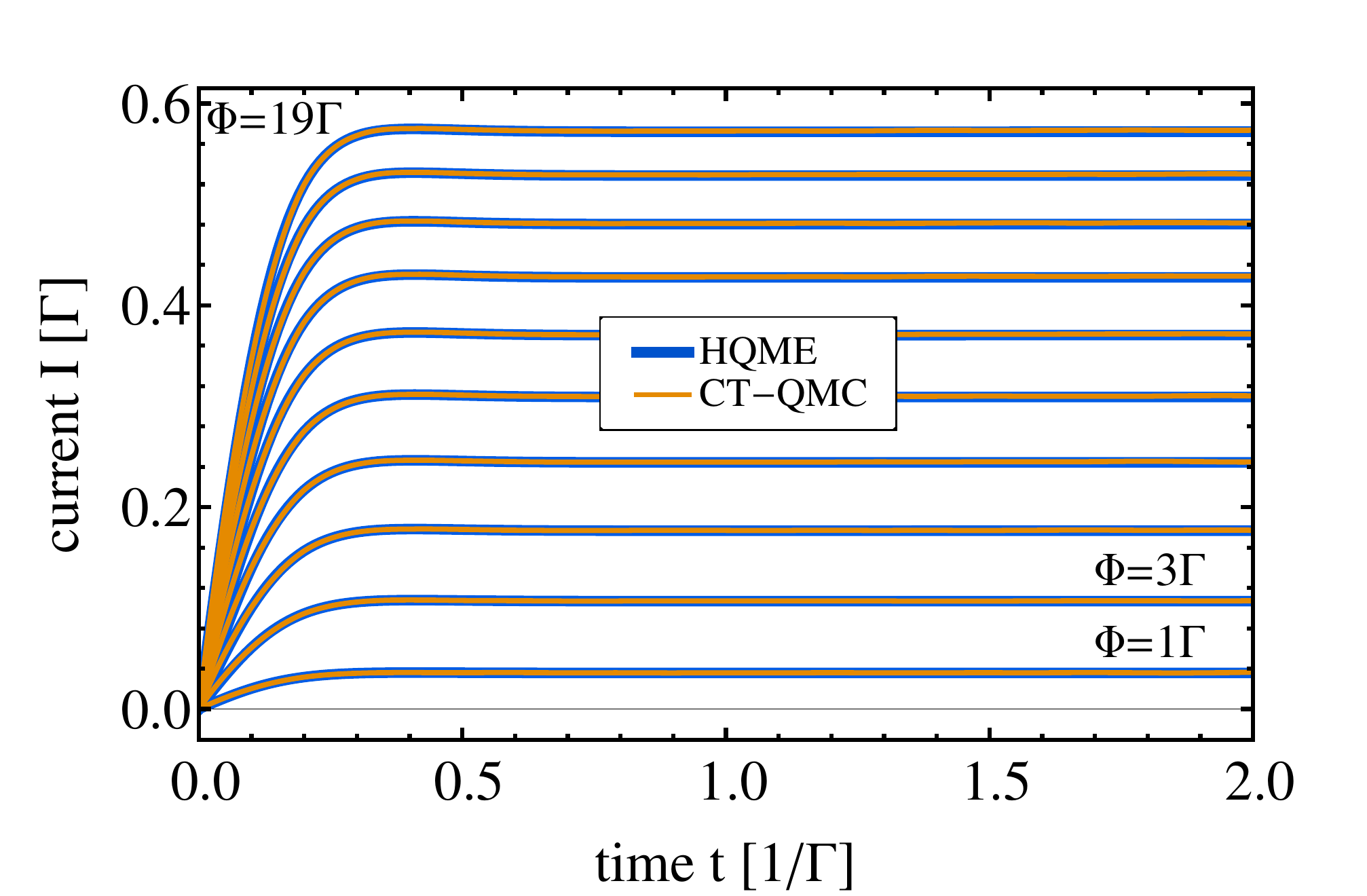}
} 
\caption{(Color online)\label{CompHigh} Symmetrized current $I=(I_{\text{L}}-I_{\text{R}})/2$
flowing through the impurity at $k_{\text{B}}T=5\Gamma$ as a function
of time $t$ for a sequence of equally spaced bias voltages $\Phi=\Gamma,3\Gamma,..,19\Gamma$ 
where the conduction bands are not shifted with the bias voltage ($S=0$). 
The model parameters used to obtain this data are summarized
in Tab.\ \ref{parameters}. After a linear increase, the current
saturates to a stationary value on a voltage-independent time scale $0.2/\Gamma$.
The HQME (blue lines) and CT-QMC methods (orange lines) give identical results to within the numerical
resolution of the data. 
}
\end{figure}

We begin with the case $S=0$, corresponding to fixed bands in the
electrodes. Fig.\ \ref{CompHigh} represents the symmetrized current
$I=(I_{\text{L}}-I_{\text{R}})/2$ flowing through the impurity as
a function of time at 
$\beta=0.2/\Gamma$. The different lines correspond to bias voltages
$e\Phi=\Gamma$, $3\Gamma$, ... $19\Gamma$. The thick blue and thin
orange lines depict results that have been obtained using the HQME
and CT-QMC scheme, respectively. The overlap between matching pairs
of lines demonstrates the agreement between HQME and CT-QMC in this
range of temperatures.

The dynamics seen in Fig.\ \ref{CompHigh} can be described and understood
as follows. Initially, for $t\lesssim0.2/\Gamma$, the current increases
almost linearly in time with a slope that is increasing linearly with the applied bias voltage. 
After this initial increase, the current level saturates rapidly to
a stationary state. The short-time behavior can be understood quantitatively 
from the relation 
\begin{eqnarray}
\frac{\text{d}}{\text{d}t}I_{K}(0) & = & 
2e\sum_{\sigma}\left(C_{K,\sigma}^{+}(0)\langle 
d_{\sigma}d_{\sigma}^{\dagger}\rangle-C_{K,\sigma}^{-}(0)\langle d_{\sigma}^{\dagger}d_{\sigma}\rangle\right),
\end{eqnarray}
which follows directly from the operator equations of motion and the choice of a product initial state. 
$C_{K,\sigma}$ is defined 
in Eq.\ (\ref{corrfunctions}). 
For an initially unoccupied impurity, the slope of the symmetrized
current is therefore given by 
\begin{eqnarray}
\frac{\text{d}}{\text{d}t}I(t) & = & 2e\sum_{\sigma}\int_{-\infty}^{\infty}\frac{\text{d}\omega}{2\pi}\left(f_{\text{L}}(\omega)\Gamma_{\text{L},\sigma}(\omega)-f_{\text{R}}(\omega)\Gamma_{\text{R},\sigma}(\omega)\right).\label{initcurrent}
\end{eqnarray}
For large band width $D$, the energy dependence of the hybridization
strengths $\Gamma_{\text{R},\sigma}(\omega)$ can be neglected. 
The slope of the current is then solely determined by the difference between 
the Fermi functions. The latter is proportional to the applied bias voltage 
$\Phi$. 
Note that the initial dynamics cannot be linked to a single time scale here but is 
influenced by the position of the energy levels, the band-width $D$ and the temperature $T$.

\begin{figure}
\resizebox{\newwidth}{\newheight}{
\includegraphics{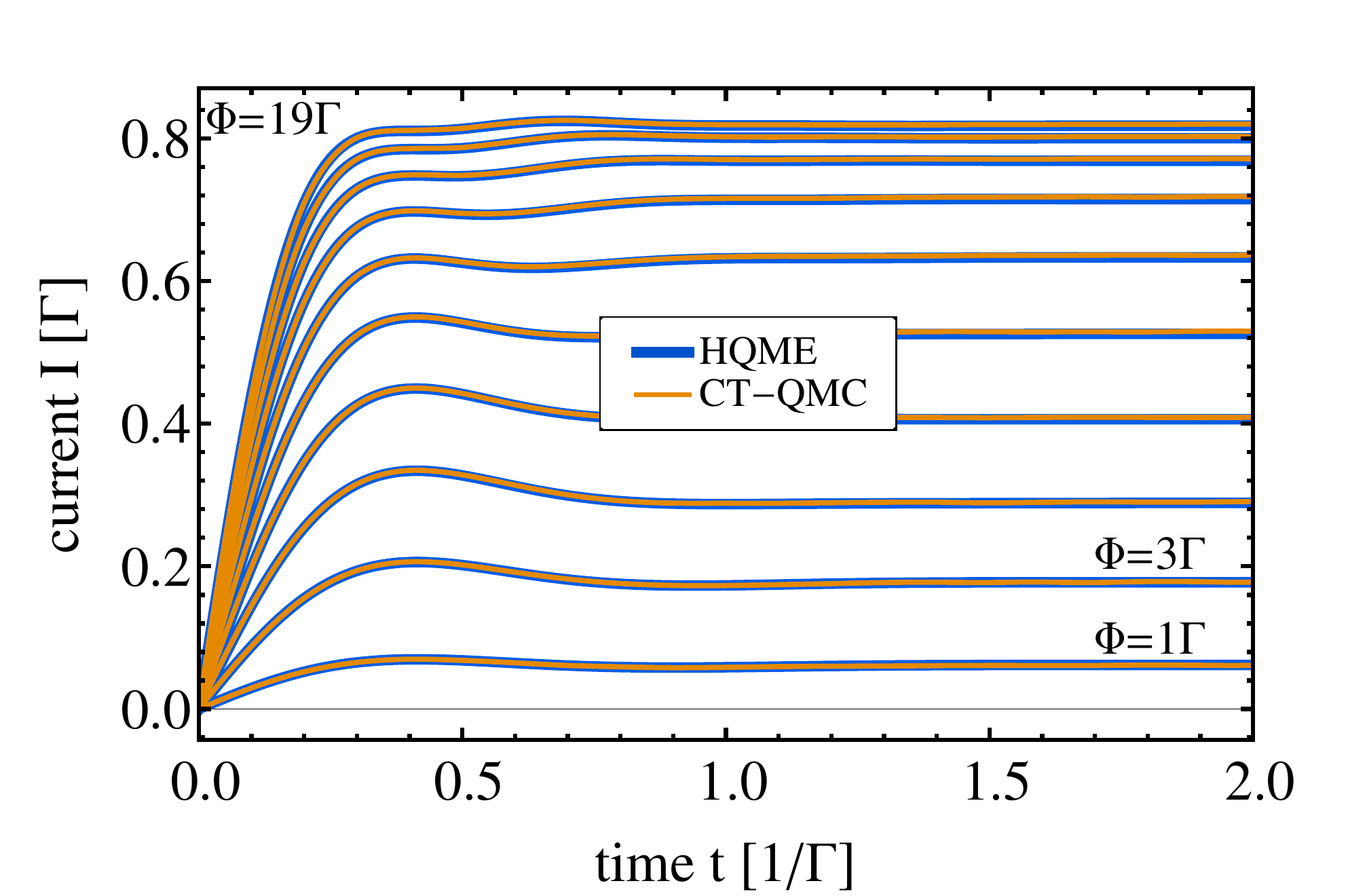}
} 
\caption{(Color online)\label{CompInter} 
Symmetrized current $I=(I_{\text{L}}-I_{\text{R}})/2$
flowing through the impurity at $k_{\text{B}}T=\Gamma$ as a function
of time $t$ for a sequence of equally spaced bias voltages $\Phi=\Gamma,3\Gamma,..,19\Gamma$ 
where the conduction bands are not shifted with the bias voltage ($S=0$). 
The model parameters used to obtain this data are summarized
in Tab.\ \ref{parameters}. After a linear increase, the current
level slightly oscillates before it reaches its stationary value.
The corresponding time scales are given by the dynamical phases of
the problem and the inverse temperature $1/(k_{\text{B}}T)=1/\Gamma$,
respectively. 
}
\end{figure}

\begin{figure}
\resizebox{\newwidth}{\newheight}{
\includegraphics{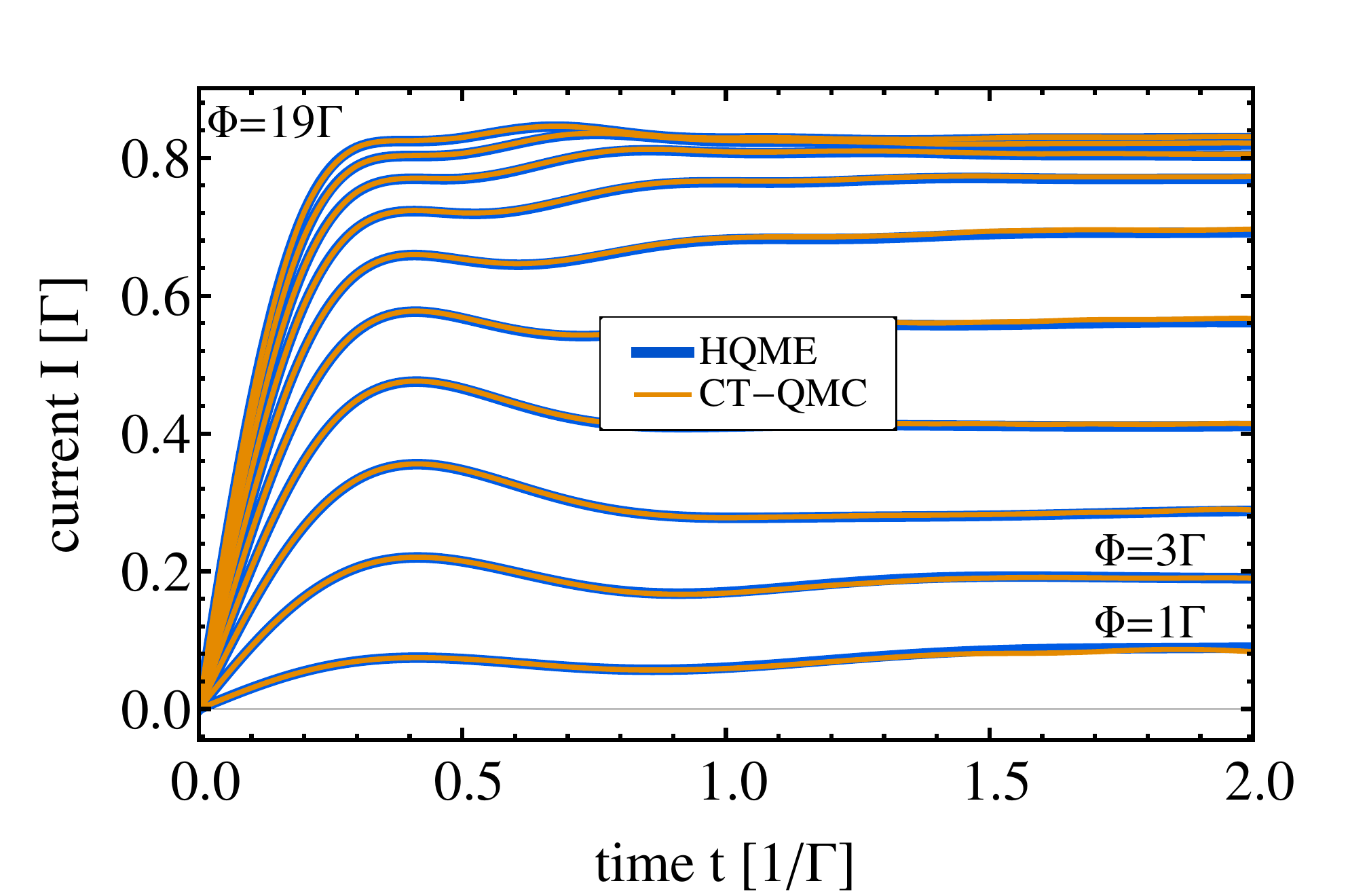}
} 
\caption{(Color online)\label{CompLow} Symmetrized current $I=(I_{\text{L}}-I_{\text{R}})/2$
flowing through the impurity at $k_{\text{B}}T=\Gamma/5$ as a function
of time $t$ for a sequence of equally spaced bias voltages $\Phi=\Gamma,3\Gamma,..,19\Gamma$ 
where the conduction bands are not shifted with the bias voltage ($S=0$). 
The model parameters used to obtain this data are summarized
in Tab.\ \ref{parameters}. The oscillations of the current level
that appear right after the initial linear increase become more pronounced at lower temperatures. 
The corresponding time scales are given by the dynamical phases of
the problem and the inverse temperature $1/(k_{\text{B}}T)=\Gamma/5$,
respectively. Deviations between the HQME (blue lines) and CT-QMC results (orange lines) are consistent.
}
\end{figure}

At lower temperatures, both schemes require a larger computational
effort in order to reach the same level of precision as compared to
higher temperatures. This is demonstated in Figs.\ \ref{CompInter}
and \ref{CompLow}, which show the time-dependent current of our setup
at lower temperatures $\beta=1/\Gamma$ and $\beta=5/\Gamma$, respectively.
The data has been obtained with a similar numerical effort as that
shown in Fig.\ \ref{CompHigh}. One observes that both schemes agree
very well, but small deviations, which are consistent with the applied accuracy, 
begin to occur. 

As the temperature $T$ decreases, 
coherent processes become more important and give rise 
to oscillations of the current level (see Figs.\ \ref{CompInter} and \ref{CompLow}). 
The period of these oscillations 
is given by the dynamical phases of the system, in particular the
difference between the energy levels of the impurity and the chemical
potentials in the electrodes. Accordingly, the dynamical oscillations
of the current level show a bias dependence, which is clearly visible
in our data. This behavior is known as \emph{current ringing} and
has been outlined before in a slightly different context, namely as
a response to bias voltage pulses and/or quenches \cite{Wingreen93,Taranko2012}.

Another bias dependence appears in the stationary values of the current
level. For high temperatures, thermal broadening leads to an almost
linear increase of the stationary current, at least in the range of
bias voltages considered here. This is evident from the almost equidistant
values in Fig.\ \ref{CompHigh}. The stationary values seen in Figs.\ \ref{CompInter}
and \ref{CompLow} are clearly non-equidistant. This indicates a strong
non-Ohmic saturation of the current level with increasing bias voltage, 
which originates from the restricted number of conductance channels through 
the impurity. 

We conclude at this point that the agreement between HQME and CT-QMC
results is very good in the parameter ranges we have studied. 
We corroborated this statement for a number of other setups, where
we changed the position of the energy levels, $\epsilon_{\uparrow/\downarrow}\neq-U/2$,
introduced a level splitting / magnetic field, $\epsilon_{\uparrow}-\epsilon_{\downarrow}\neq0$,
a different shift of the conduction bands, $S=1$ (see below), and
different band widths (data not shown). Changing the electron-electron interaction strength, 
we also observed that the results converge faster for lower values of $U$ 
(which is consistent with the fact that, for $U=0$, the hierarchy (\ref{hierarcheom}) 
terminates at the second tier \cite{Jin2008,Jin2010}). 
At lower temperatures, $\beta\gtrsim\beta_{\text{Kondo}}\approx15/\Gamma$,
we were not able to converge the HQME expansion to a satisfactory
level. 

\begin{figure}
\resizebox{\newwidth}{\newheight}{
\includegraphics{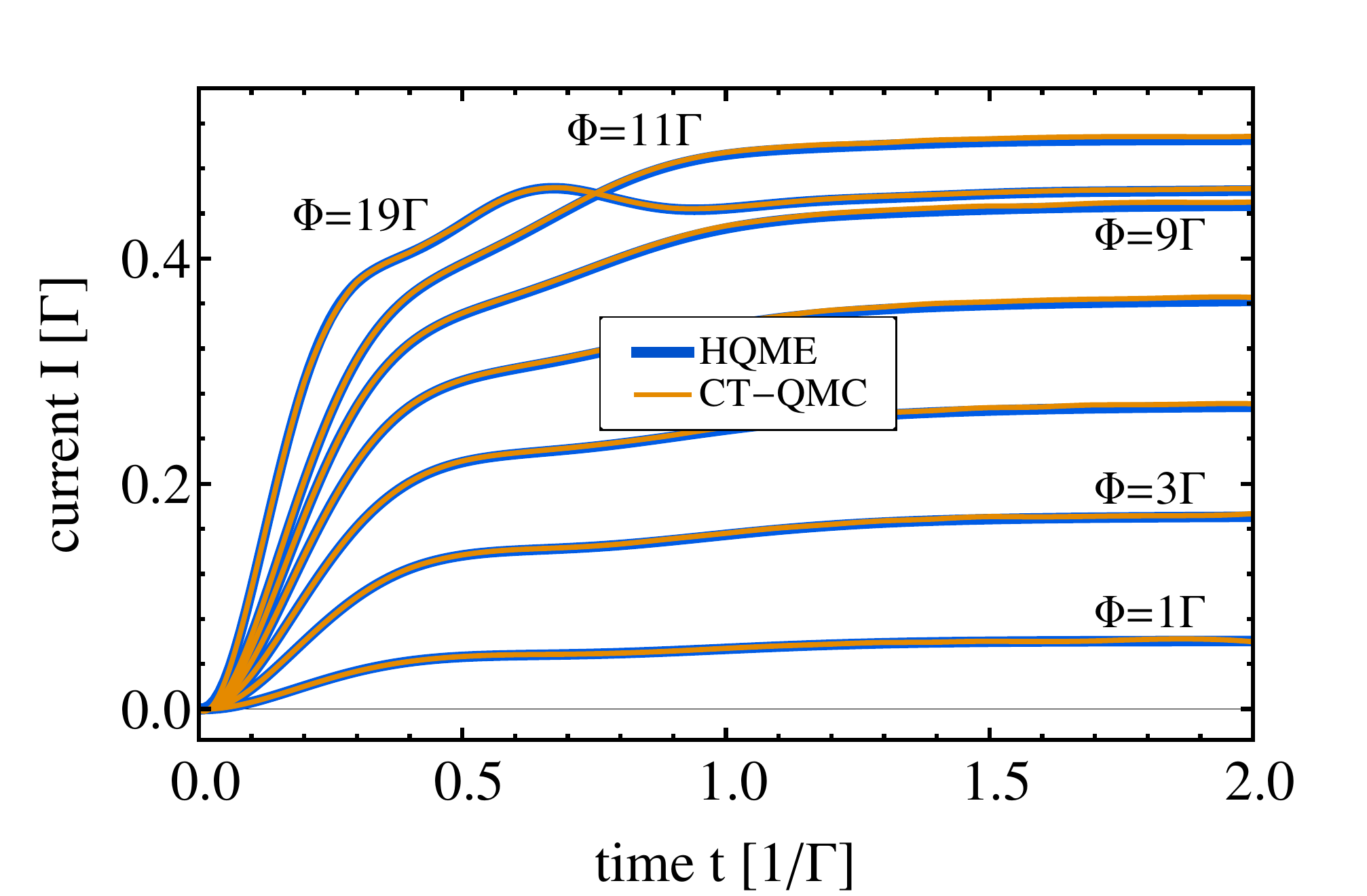}
} 
\caption{(Color online)\label{CompShift} 
Symmetrized current $I=(I_{\text{L}}-I_{\text{R}})/2$
flowing through the impurity at $k_{\text{B}}T=\Gamma$ as a function
of time $t$ for a sequence of equally spaced bias voltages 
$\Phi=\Gamma,3\Gamma,..,11\Gamma$ and $\Phi=19\Gamma$ 
where the conduction bands are shifted with the bias voltage ($S=1$). 
The model parameters used to obtain this data are summarized
in Tab.\ \ref{parameters}. In contrast to Figs.\ \ref{CompHigh}
-- \ref{CompLow}, the current level is initially not linearly increasing with the 
time $t$ when the conduction bands are shifted with the applied bias voltage
($S=1$).
}
\end{figure}

Next, we consider electrodes in which the conduction bands are shifted
with the applied bias voltage ($S=1$). The time-dependent current
of such a system is shown in Fig.\ \ref{CompShift}. It should be
compared and contrasted with the data shown in Fig.\ \ref{CompInter}.
Two qualitative differences are apparent: the first is that the slope
of the current vanishes at $t=0$. This can be easily understood from Eq.\ (\ref{initcurrent}),
as the integral on the right-hand side vanishes for shifted conduction
bands. The latter is no longer true for times $t>0$, where, initially, an increase of the current $\sim t^3$ 
is inherited from a change of the populations $\sim t^2$ \cite{Thoss01,Egorova03,Hartle2014}. 
After this initial phase, the behavior of the unshifted bands is recovered. 
The other difference is the reduced 
current level for $e\Phi=19\Gamma$, which falls below the line for
$e\Phi=11\Gamma$ at times $t>0.8/\Gamma$. This negative differential
resistance originates from both the shift of the conduction bands
and their finite band width $D$. This behavior is also well known, for example, from
transport through resonant tunneling diodes \cite{Davies93,Hyldgaard94}.

\subsection{Evolution of the magnetization for a symmetric and 
an asymmetric coupling to the electrodes}
\label{magsec}

The HQME method is particularly promising because it allows us to
calculate the exact time evolution of a correlated many-body system
with a numerical effort that scales linearly with the simulation time.
In order to demonstrate this, we discuss the magnetization, as introduced
in Eqs.\ (\ref{effectofmagneticfield}). Other observables, like populations or
the current, approach their steady state values much faster and are,
therefore, less suitable for the present purpose. In addition, we
consider an asymmetric coupling to the electrodes. As we will see,
the corresponding dynamics shows a richer set of behaviors than in
the symmetrically coupled case, and occurs on significantly longer time scales.

\begin{figure}
\begin{tabular}{ccc}
\includegraphics[scale=0.25]{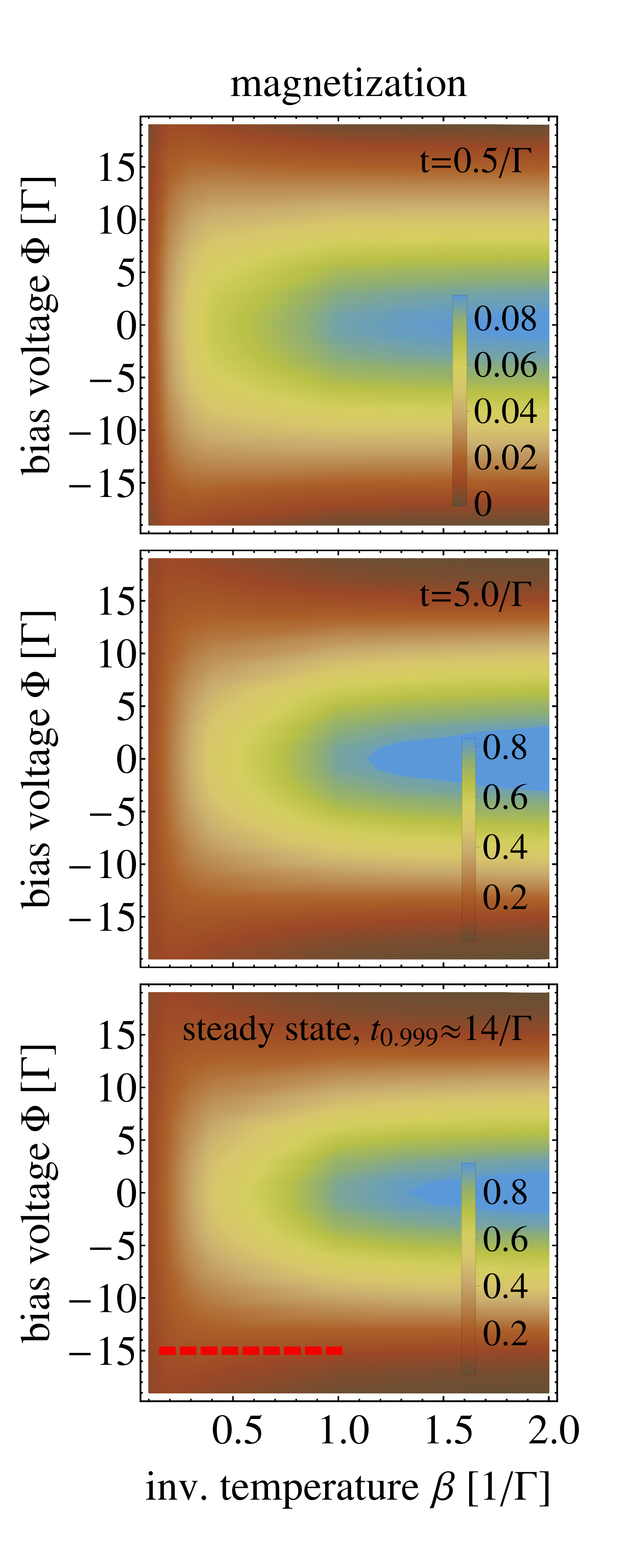} 
&
\includegraphics[scale=0.25]{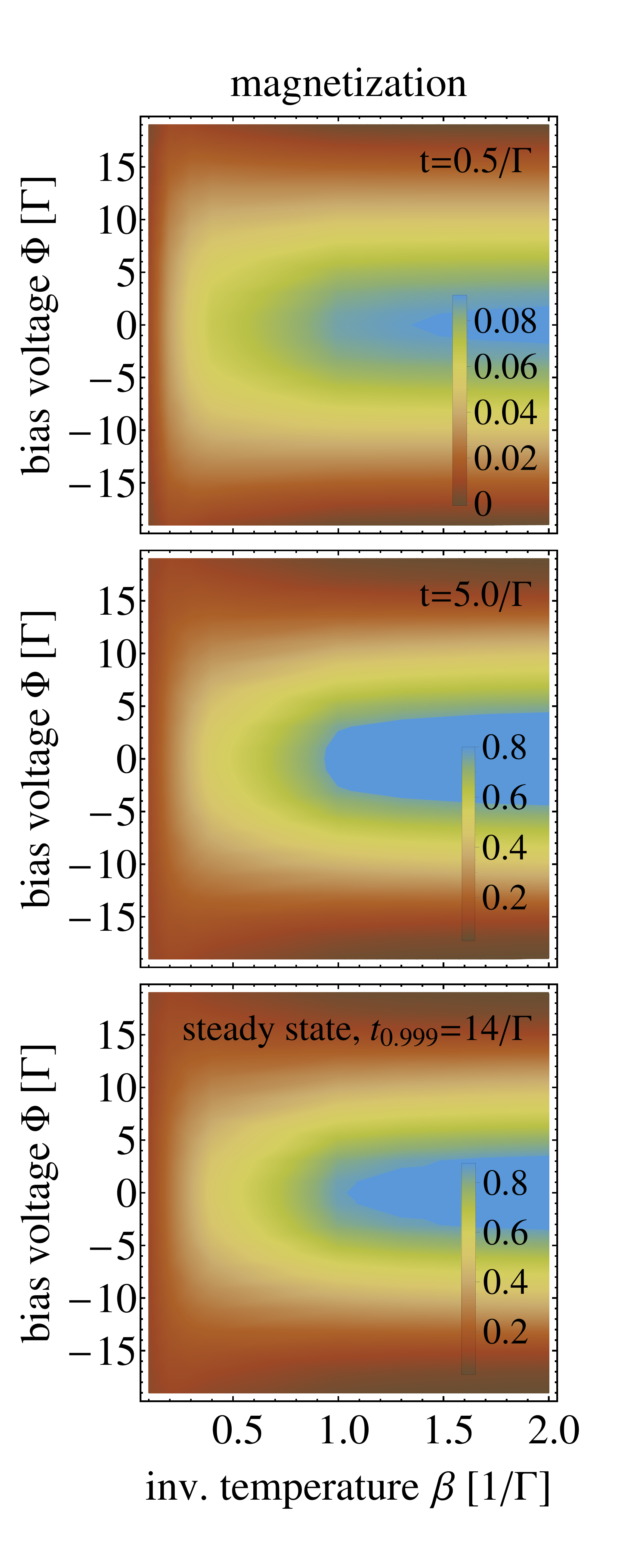} 
&
\includegraphics[scale=0.25]{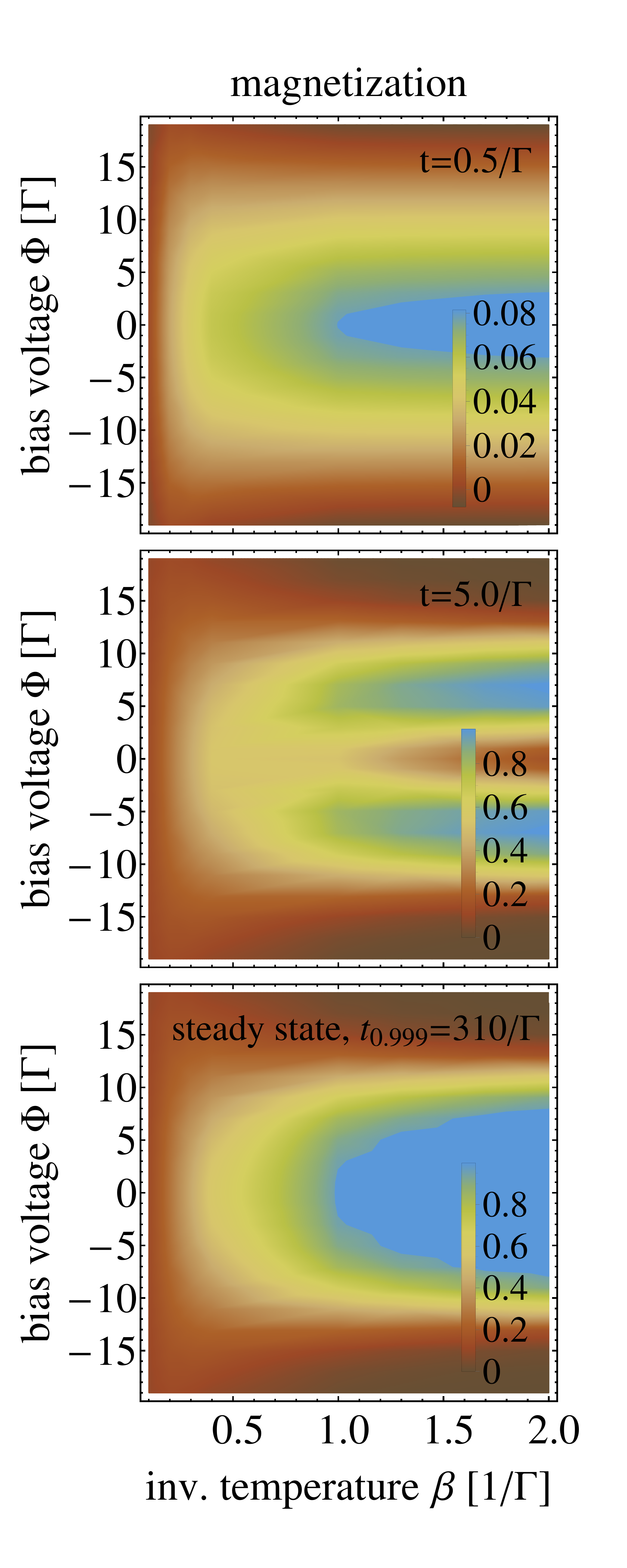} 
\end{tabular}
\caption{(Color online) \label{SymmB} Magnetization of an Anderson impurity
that is symmetrically coupled to the electrodes as a function of temperature
and bias voltage ($S=0$), and for three different times (top row: $t=0.5/\Gamma$;
middle row: $t=5/\Gamma$; bottom row: steady state). The left column
depicts the full HQME result. The middle and the right column has
been obtained by truncating the hierarchy (\ref{hierarcheom}) at
the second and the first tier, respectively. The magnetization along the red dashed line in the lower left plot 
is also depicted in Fig.\ \ref{mtfig}. 
}
\end{figure}

We start our analysis with a symmetrically coupled impurity in a 
strong magnetic field $h=2\Gamma$ (the behavior at other field strengths
is similar - data not shown). The corresponding magnetization
of the impurity is depicted in Fig.\ \ref{SymmB} as a function of
both bias voltage $\Phi$ and inverse temperature in the electrodes
$\beta$. We show a $3\times3$ array of plots, where the top row
represents the magnetization at short time scales, $t=0.5/\Gamma$, 
and the middle row at intermediate time scales, $t=5/\Gamma$. The
bottom row depicts the steady state values. In the latter, we also
give the time $t_{0.999}$ at which the impurity reached 99.9\%
of its final magnetization (note that this time scale is longest for
low temperatures and bias voltages). The different columns are obtained
using different levels of approximation. The left column is computed
using the full HQME approach. The second and third column are obtained
by truncating the hierarchy (\ref{hierarcheom}) at the second and
the first tier. This corresponds to a hybridization
expansion to $\mathcal{O}((\Gamma/(k_{\text{B}}T))^{2})$ and $\mathcal{O}(\Gamma/(k_{\text{B}}T))$,
respectively (see Sec.\ \ref{hqmesec}). The different levels of
approximation facilitate a discussion of the relevant processes and
mechanisms, for example, where and when processes of higher order 
are important (see below).

The data of Fig.\ \ref{SymmB} can be understood as follows. As we
start from a state with zero magnetization, the magnetization at short
time scales is almost an order of magnitude smaller than the final
steady state values. The maximum magnetization is obtained at small bias voltages
and temperatures. At higher temperatures and voltages, where the impurity
exchanges particles with the electrodes at a wider range of energies,
the magnetization becomes quenched. We would like to emphasize that
the system reaches its steady state not before times $t\gtrsim15/\Gamma$.

\begin{figure}
\resizebox{\newwidth}{\newheight}{
\includegraphics{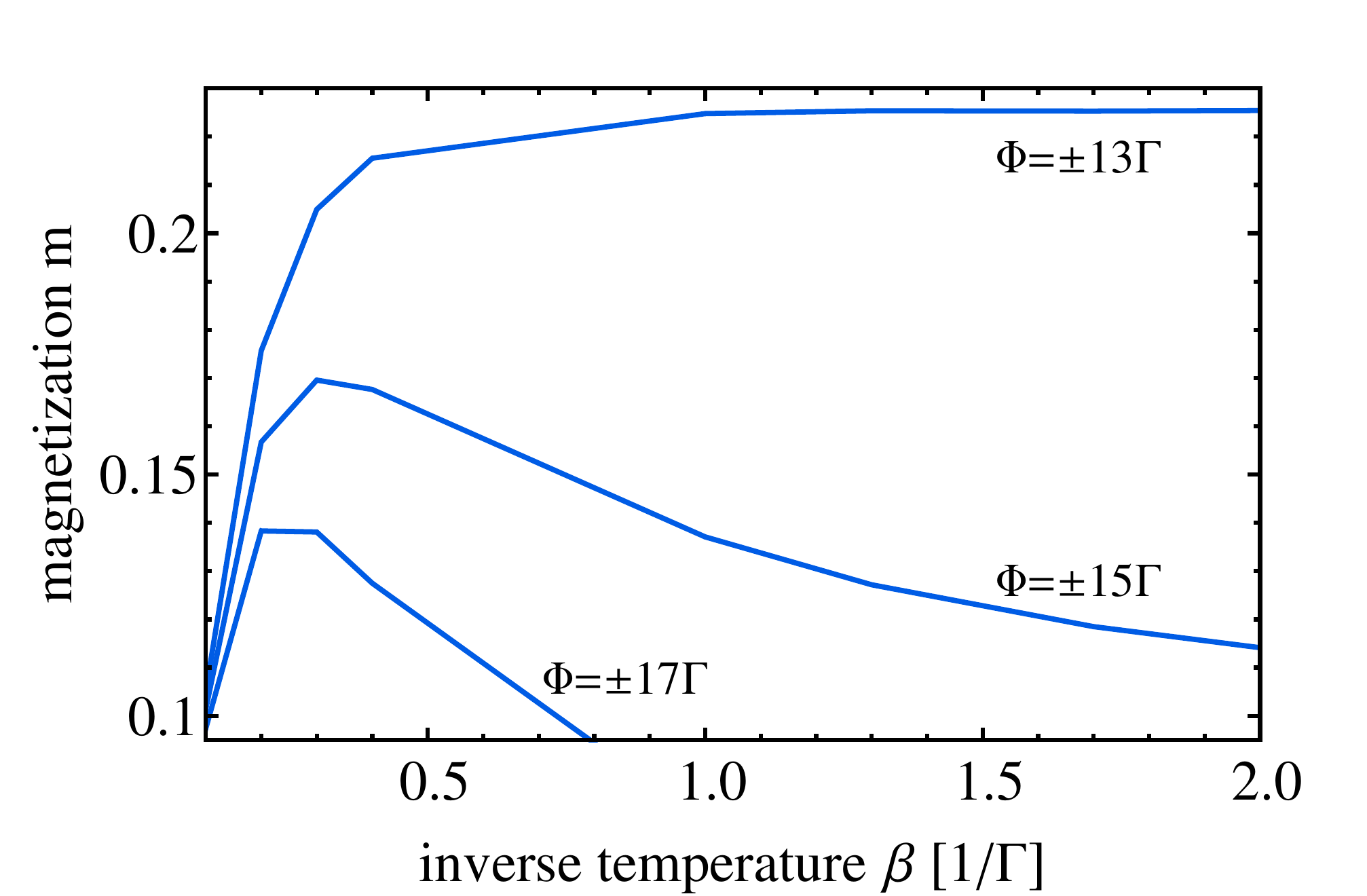}
}
\caption{(Color online)\label{mtfig} 
Magnetization of an Anderson impurity
that is symmetrically coupled to the electrodes as a function of temperature 
in the steady state and applied bias voltages $\Phi=\pm13\Gamma$, $\pm15\Gamma$  
and $\pm17\Gamma$  ($S=0$). Note that the kinks originate from a simple 
linear interpolation between the data points. 
}
\end{figure}

The exact result is very similar to the one that is obtained with
a second order truncation of the hierarchy (\ref{hierarcheom}). A
tendency towards a higher magnetization is visible if the number of
exchange processes that is taken into account in our calculations
is reduced. Truncation at the first tier enhances the effect, but
also results in a qualitatively different structure of the magnetization
at intermediate time scales (right plot of the middle row). Here,
the magnetization shows a peak at positive and negative voltages,
while the exact and second order result exhibit only a single peak that is centered 
around zero bias. 

The splitting of the peak magnetization can be understood with the
bias dependence of resonant exchange processes with the electrodes.
At zero bias, the difference between the chemical potentials in the
electrodes and the single-particle levels of the impurity (see Fig.\ \ref{qdfig}b)
is largest and resonant processes are strongly suppressed. This suppression
is less pronounced at non-zero voltages such that the steady state magnetization
can develop on shorter time scales. Accordingly, this behavior shows
only a weak temperature dependence, resulting in an almost horizontal
splitting of the peak that is seen in Fig.\ \ref{SymmB}. This splitting
is not seen at short times scales (right plot of the top row), because
the initial state is not decaying exponentially at short times. It
rather shows a power-law decay, $\sim t^{2}$, which is well known
from an analysis of similar systems in terms of Born-Markov theory
\cite{Thoss01,Egorova03,Hartle2014}.

Another intriguing effect occurs at higher bias voltages ($\vert e\Phi\vert>10\Gamma$).
Here, the magnetization shows a non-monotonic behavior with respect
to temperature: it becomes stabilized by increasing temperature before
decreasing again at temperatures $\beta\lesssim0.2/\Gamma$ (follow,
e.g., the red dashed line in Fig.\ \ref{SymmB} from $\beta=1/\Gamma$
to $\beta=0.1/\Gamma$). This non-monotonic behavior is explicitly depicted 
in Fig.\ \ref{mtfig}, where the magnetization $m$ is shown, for example,  
as a function of the inverse temperature $\beta$ and fixed bias voltages $\Phi=\pm13\Gamma$, 
$\pm15\Gamma$ and $\pm17\Gamma$. 
This non-linear dependence of the magnetization on temperature 
was discovered only recently (see Ref.\ \cite{Cohen2013}). It is most pronounced
in the steady state and for a truncation of the hierarchy (\ref{hierarcheom})
at a lower tier. The latter points towards a physical interpretation
of the effect, suggesting that it originates from the broadening of
the peak magnetization around zero bias with increasing temperature and 
the quenching of the magnetization at very high temperatures.

\begin{figure}
\begin{tabular}{ccc}
\includegraphics[scale=0.25]{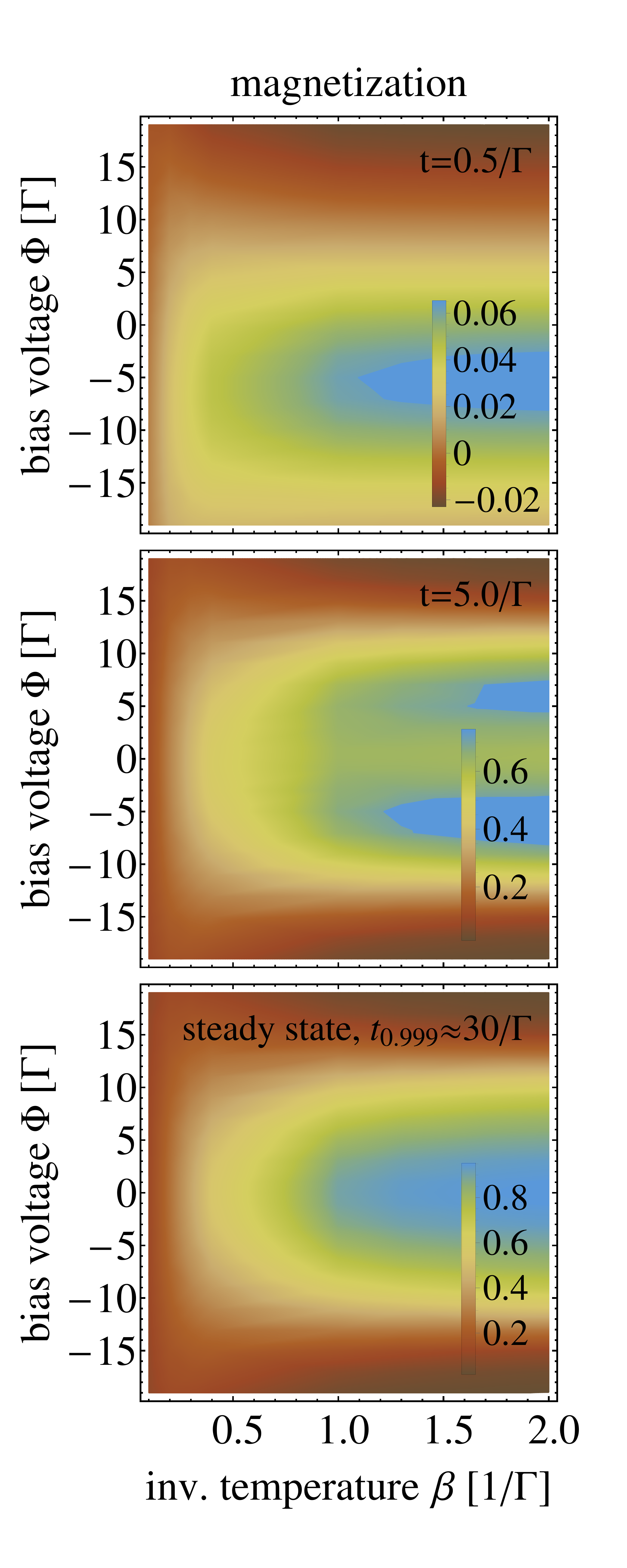} 
&
\includegraphics[scale=0.25]{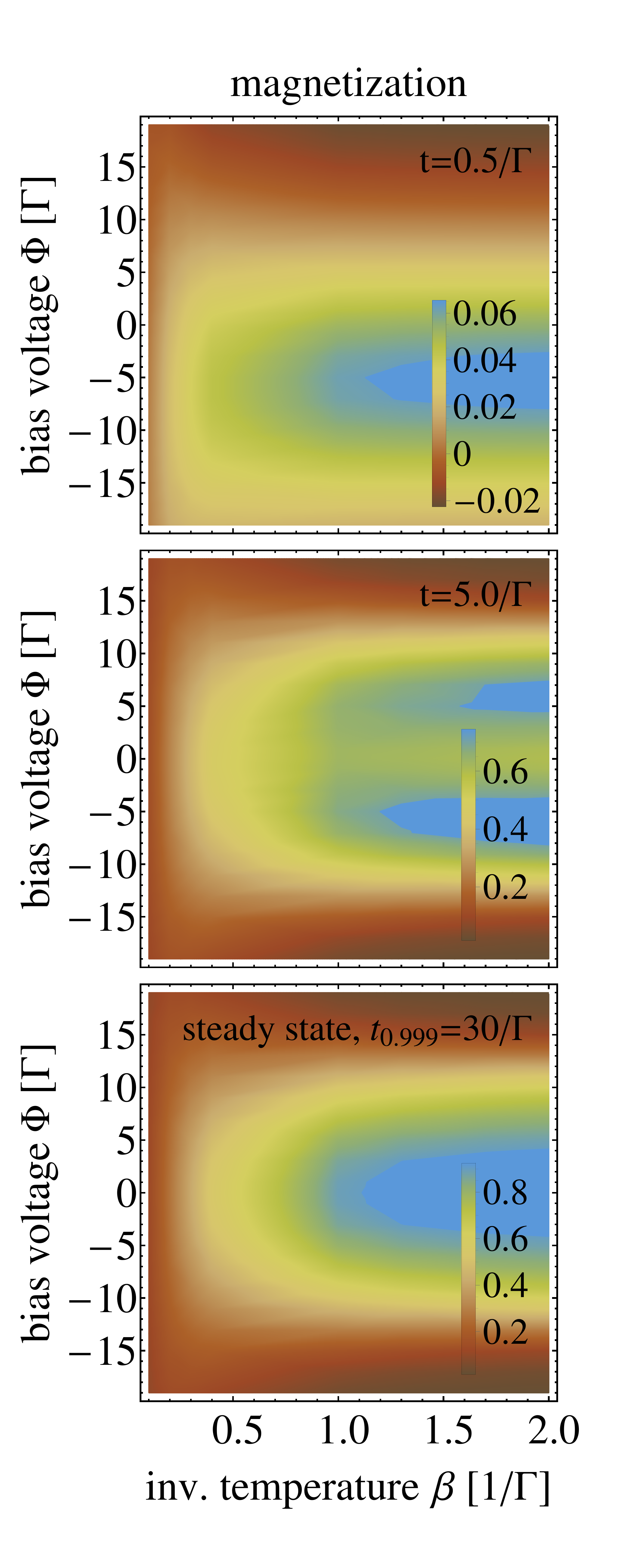} 
&
\includegraphics[scale=0.25]{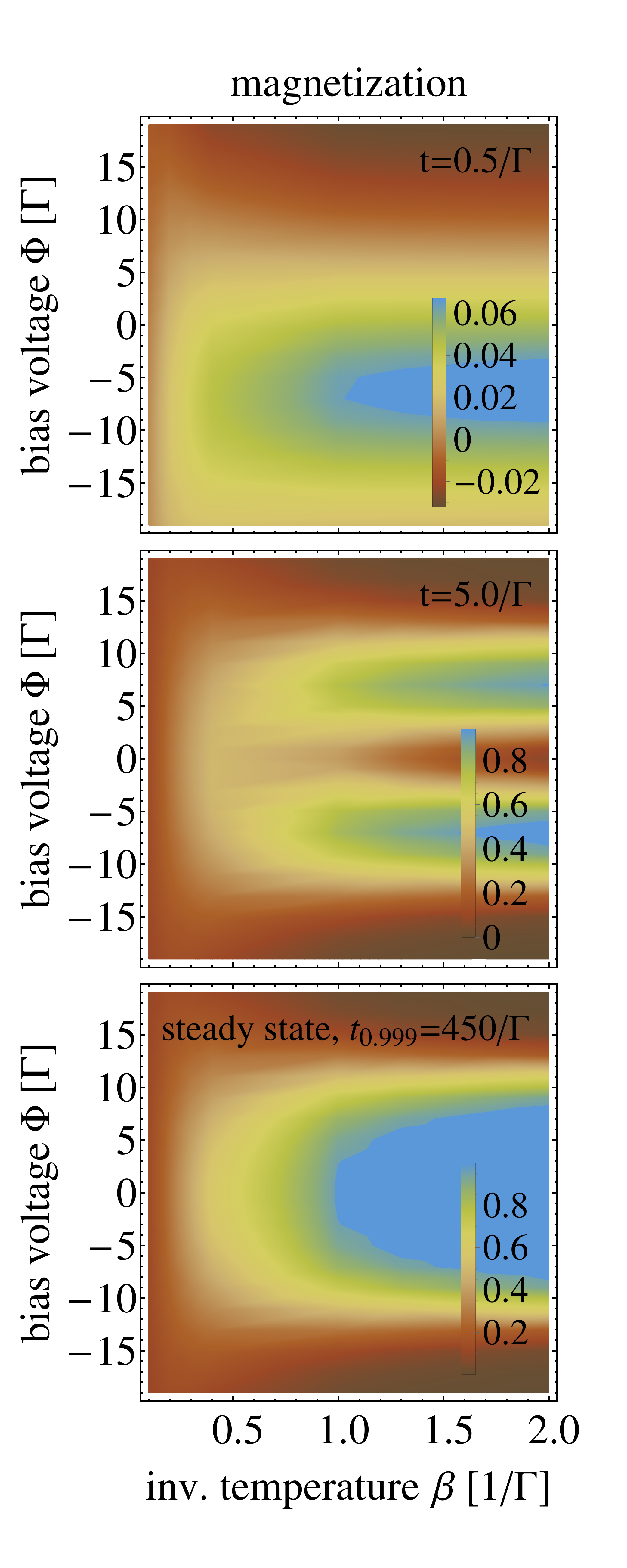} 
\end{tabular}
\caption{(Color online) \label{ASymmB} 
Magnetization of an Anderson impurity
asymmetrically coupled to the electrodes as a function temperature
and bias voltage ($S=0$) and for three different times (top row: $t=0.5/\Gamma$;
middle row: $t=5/\Gamma$; bottom row: steady state). The left column
depicts the full HQME result. The middle and the right column has
been obtained by truncating the hierarchy (\ref{hierarcheom}) at
the second and the first tier, respectively. 
}
\end{figure}

The last scenario we discuss is an asymmetric coupling of the impurity
to the electrodes. The corresponding magnetization is shown in Fig.\ \ref{ASymmB}.
It can be directly compared to the magnetization of the symmetrically
coupled impurity that is depicted in Fig.\ \ref{SymmB}. It can be
seen that, on short time scales (top row), the initial peak of the
magnetization is shifted towards negative voltages. This is because
the coupling to the right electrode is weaker and we start with an
initially unoccupied system. The initial population of the impurity
is therefore dominated by exchange processes with respect to the left
electrode. The corresponding dynamics occurs on shorter time scales 
for negative voltages, because the chemical potential of the left electrode $\mu_{\text{L}}$
is then closer to the single-particle levels $\epsilon_{\uparrow/\downarrow}$.

Another qualitative difference with respect to the symmetric case
occurs on intermediate time scales. Here, the magnetization is peaked
at non-zero values of the bias voltage $\Phi$ even when higher-order 
processes are taken into account. This behavior was also
seen in the symmetric case but only if the hierarchy (\ref{hierarcheom})
is truncated at the first tier, that is by disregarding higher-order
processes. These processes become quenched by the weaker coupling
to the right electrode, while resonant processes with respect to the
left electrode are not. The situation is therefore similar to the
symmetrically coupled case without higher-order processes. 
The magnetization of the impurity evolves to very similar steady state
values, but on even longer times scales ($t\gtrsim30/\Gamma$). Minor
differences occur due to the weaker hybridization with the right electrode
(\emph{i.e.} a less pronounced broadening of the energy levels).

We close this section by a discussion on the generality of our findings.
We observed, for example, a very similar behavior of the magnetization
dynamics for different choices of the voltage division factor ($\mu_{\text{L}}\neq-\mu_{\text{R}}$
for $\Phi\neq0$, data not shown). We also calculated the magnetization
dynamics starting from different initial states. If, for example,
the initial magnetization points in the direction of the magnetic
field, the impurity magnetization shows a very similar behavior as
compared to the one we discussed for a symmetric coupling to the electrodes.
Similar structures as for an asymmetric coupling to the electrodes
appear, if the initial magnetization points opposite to the applied
magnetic field. These signatures, however, decay on much shorter time
scales due to the stronger coupling to the right electrode.

\section{Conclusion}

In this work, we give the first direct comparison of the hierarchical
quantum master equation method \cite{Tanimura2006,Welack2006,Jin2008,Hartle2013b,Hartle2014}
and the diagrammatic, continuous-time quantum Monte Carlo approach
\cite{Werner2006,Werner2009,Schmidt2009,Schiro2009,Cohen2011,Cohen2013}.
To this end, we have studied the nonequilibrium transport properties
of an Anderson impurity that is coupled to two electrodes with different
chemical potentials. This transport problem represents a well established
and fairly well understood test case. We discussed the main characteristics
of the two numerically exact methods (cf.\ Sec.\ \ref{hqmevsctqmc}). 
They are distinguished by the range of parameters where exact results can be obtained in practical
calculations. CT-QMC gives access to the short- and intermediate-time
dynamics ($\lesssim$ 10 units of the inverse hybridization strength) but in
general fails to describe long-time dynamics, for example in the presence
of Kondo correlations, because the numerical effort scales exponentially
with the simulation time \cite{Cohen2013}. In contrast, the hierarchical
master equation method scales linearly with the simulation time and
is, therefore, not limited in terms of the accessible time scale.
It represents a hybridization expansion, which can be carried out
to sufficiently high order if the temperature in the electrodes is not too low. 
For the present problem, we were able to obtain converged 
results only for temperatures above the Kondo temperature. 
Our findings provide a wealth of numerically exact benchmark data certain to be useful to future
method developers. 

We have also elucidated interesting physical phenomena for the range
of parameters, where numerically exact results have been accessible
with a reasonable numerical effort. We investigated the short-time
dynamics of the (symmetrized) current flowing through the impurity
in the presence of a bias voltage, starting from a product initial
state where the impurity is not populated by electrons. While a linear
increase of the current level is found for situations where the conduction
bands are not shifted with the applied bias voltage (corresponding
to realizations of quantum dots with semiconductor heterostructures),
a qualitatively different behavior emerges when the electrodes are
not charged when applying a bias voltage. At low temperatures, oscillations
of the current level, current ringing \cite{Wingreen93}, due to dynamical phases appear. 
In addition to the current,
we also studied the magnetization dynamics in the presence of a magnetic
field. We confirmed the recently reported non-monotonic temperature
dependence of the steady state magnetization \cite{Cohen2013} and
traced it back to competing broadening effects of the impurity levels. In addition, we found 
complex structures on intermediate yet long time scales (\emph{i.e.}, in the cases 
studied, about 10 units of the inverse hybridization strength) 
in the presence of an asymmetric coupling to the electrodes (cf.\ Fig.\ \ref{ASymmB}). 

Our comparative study is a first step towards practical guidelines
in choosing the right solver for a given impurity problem. Since the
presented methods cannot cover the full spectrum of problems, it would
be interesting to compare them to other exact schemes, including,
for example, numerical renormalization group, density matrix renormalization
group, multi-layer multi-configurational time-dependent Hartree, or
other quantum Monte Carlo schemes. A primary goal is to identify regions
of parameter space where methods overlap and, of course, regions which
cannot be covered satisfactorily by any of the available methods.
Further activities in this direction are planned, and we would like
to encourage other researchers to participate in these efforts.

\section*{Acknowledgement}

We thank T.\ Pruschke for helpful discussions. 
AJM acknowledges support from DOE grant number DE-SC0012592 for his work on this project. 
RH gratefully acknowledges financial support of the Alexander von Humboldt foundation via a Feodor
Lynen research fellowship and the Deutsche Forschungsgemeinschaft (DFG) under grant No.\ HA 7380/1-1. 
GC is grateful to the Yad Hanadiv--Rothschild
Foundation for the award of a Rothschild Postdoctoral Fellowship.
We also thank the GWDG G\"ottingen for generous allocation of computing
time.

\appendix

\section{Convergence properties of HQME}
\label{convanalysis}

\begin{figure}
\begin{tabular}{l}
\hspace{-0.5cm}(a) \\
\resizebox{\newwidth}{\newheight}{
\includegraphics{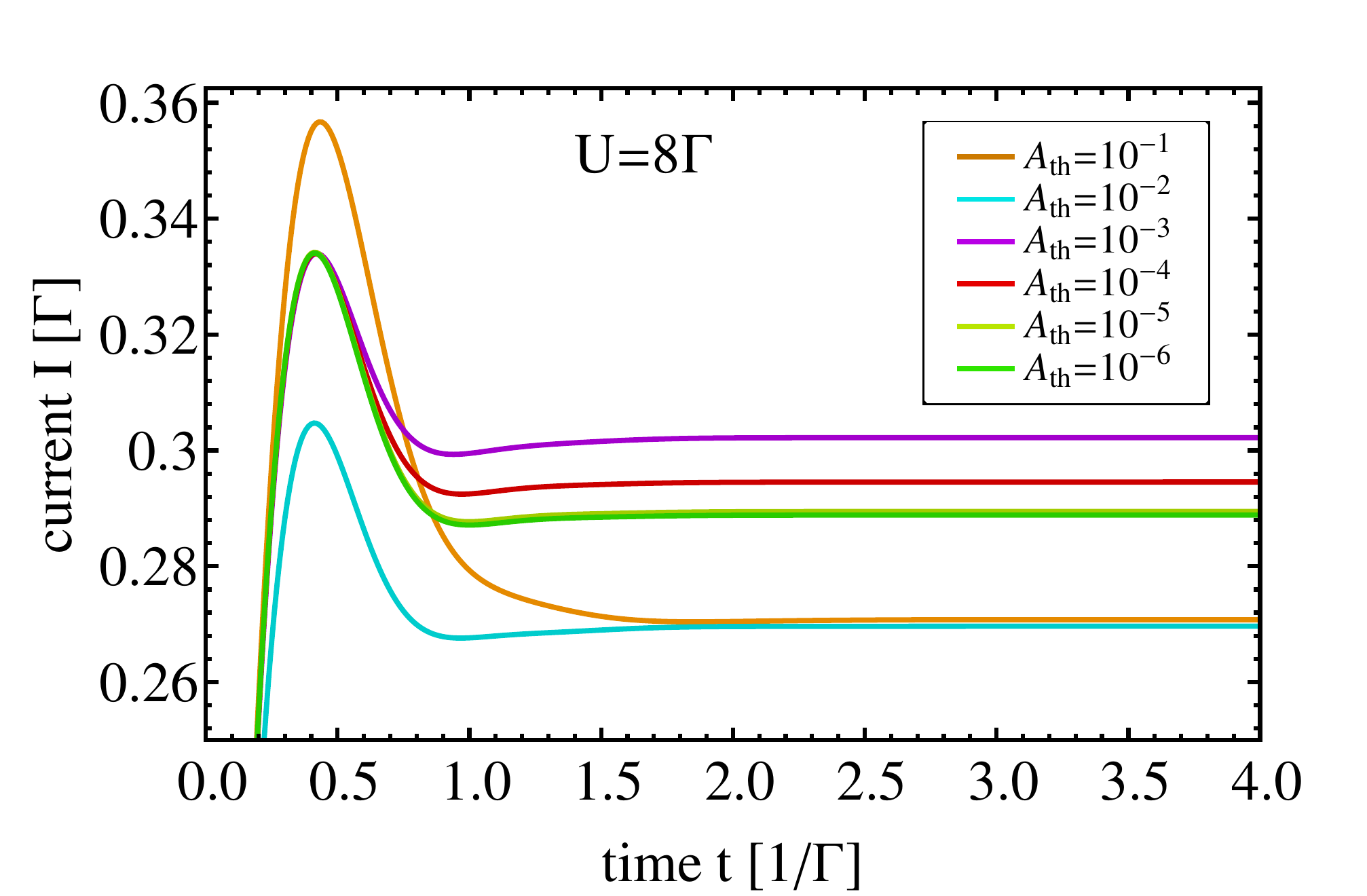}
}\\
\hspace{-0.5cm}(b) \\
\resizebox{\newwidth}{\newheight}{
\includegraphics{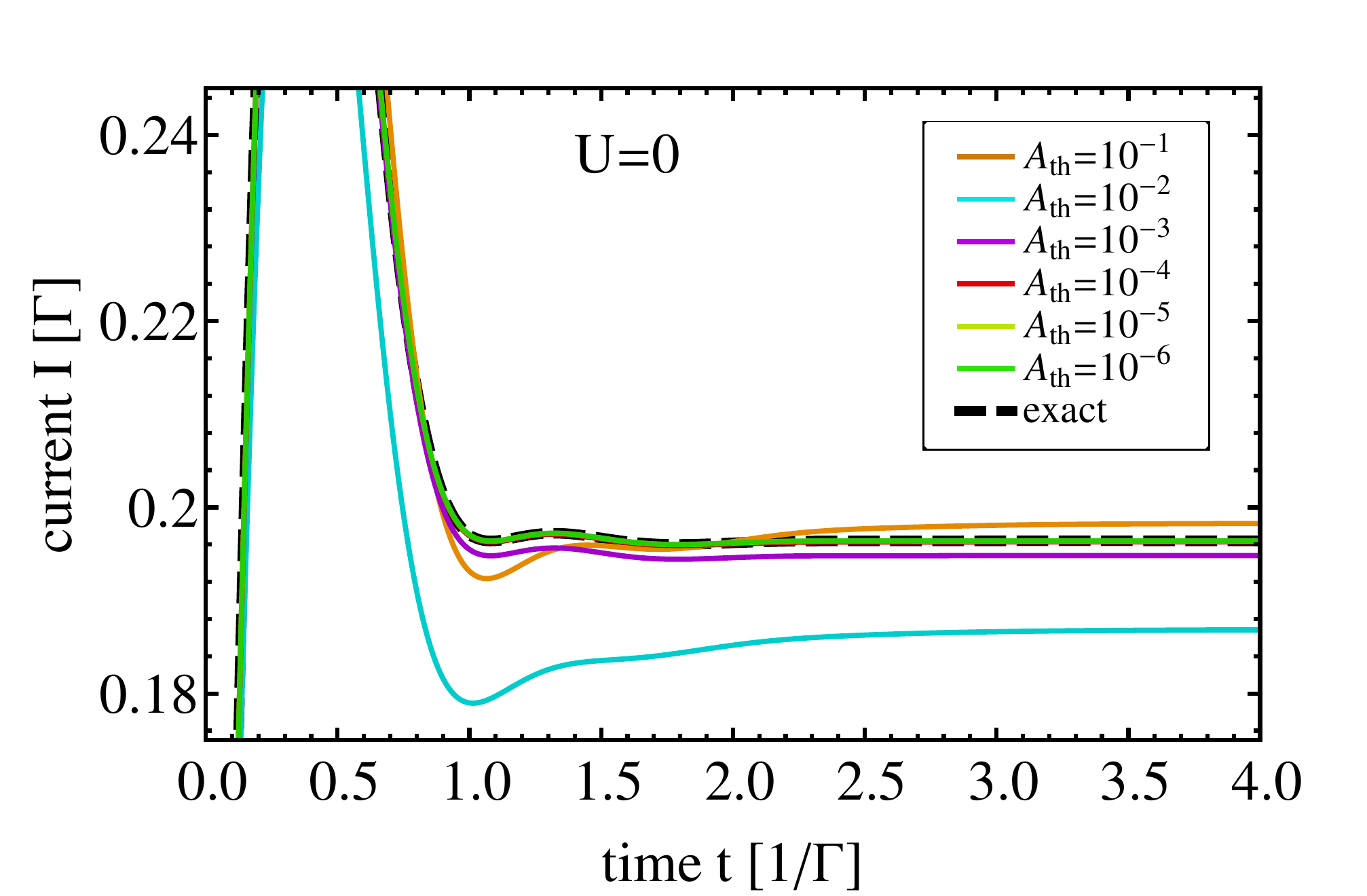}
}\\
\end{tabular}
\caption{(Color online) \label{convfig} 
Convergence analysis of the symmetrized
current $I=(I_{\text{L}}-I_{\text{R}})/2$ that is flowing through
the impurity at $k_{\text{B}}T=\Gamma$ and $\Phi=5\Gamma$ ($S=0$) as a function
of time $t$. Panel (a) shows the convergence behavior for $U=8\Gamma$.
The behavior for $U=0$ is shown in Panel (b). The model parameters
used to obtain this data are summarized in Tab.\ \ref{parameters}.
}
\end{figure}

The convergence properties of the HQME approach strongly depend on
the importance criterion that is used to truncate the hierarchy of
equations of motion (\ref{hierarcheom}). In this appendix, we demonstrate
the convergence behavior of our HQME scheme explicitly. To this end,
we reconsider the time-dependent (converged) current shown
in Fig.\ \ref{CompInter}. We replot the result for $\Phi=5\Gamma$
in Fig.\ \ref{convfig}a. It corresponds to the graph with the threshold
value $A_{\text{th}}=10^{-6}$. In addition, we plot results that
were obtained for higher threshold values $A_{\text{th}}$. In Tab.\ \ref{convtab},
we list the number of auxiliary operators (AO) that were taken into
account and the maximum tier level. The results are considered to
be converged once the threshold value is below $10^{-5}$. The corresponding 
number of AOs is $\sim10^{4}$. The respective maximum tier
level is $5$.

\begin{table}
\begin{center}
\begin{tabular}{|c|*{7}{ccc|}}
\hline 
threshold value && $10^{-1}$ &&& $10^{-2}$ &&& $10^{-3}$ &&& $10^{-4}$ &&& $10^{-5}$ &&& $10^{-6}$ &\\ \hline 
\# of AOs && 269  &&& 447  &&& 1158  &&& 3009 &&& 8317 &&& 20912 &\\
max.\ tier level && 2  &&& 3  &&& 3  &&& 4 &&& 5 &&& 5 &\\
\hline 
\end{tabular}
\end{center}
\caption{\label{convtab} Number of auxiliary operators for different threshold values 
of the importance criterion (\ref{ampli}). 
}
\end{table}

In Sec.\ \ref{Comparison}, we have shown that our (converged) results
coincide with the ones obtained from CT-QMC. This demonstrates the
validity and usefulness of our truncation scheme. At this point, we
would like to give an additional proof of this statement by showing
the convergence of our scheme towards the solution of an analytically
solvable case. Fig.\ \ref{convfig}b has been obtained with the same
parameters as Fig.\ \ref{convfig}b, except that we 'turned off'
electron-electron interactions, \emph{i.e.}\ we used $U=0$. In this
limit, the exact result is known and can be obtained, for example,
by truncating the hierarchy (\ref{hierarcheom}) at the second tier
(without applying the importance criterion (\ref{ampli})) \cite{Jin2008,Jin2010}.
It can be seen that our results converge to the exact result and that
convergence is faster than in the interacting case.

\end{document}